\documentclass[aps,pre]{revtex4}
\usepackage{epsfig,graphics,amssymb,amsmath,subeqnarray,color}

\def \bsigma{\boldsymbol{\sigma}}
\def \bsigmat{\tilde{\boldsymbol{\sigma}}}
\def \bu{\mathbf{u}}
\def \hb{\bar{h}}

\def \but{\tilde{\mathbf{u}}}

\def \UD{U_\Delta}
\def \UDone{U_{\Delta 1}}
\def \UDtwo{U_{\Delta 2}}
\def \UDthree{U_{\Delta 3}}
\def \UDfour{U_{\Delta 4}}
\def \UDm{U_{\Delta m}}
\def \SR{\dot{\text{S}}_{\text{R}}}
\def \d{\text{d}}
\def \bG{\mathbf{G}}
\def \be{\mathbf{e}}
\def \bT{\mathbf{T}}
\def \bx{\mathbf{x}}
\def \bf{\mathbf{f}}
\def \bn{\mathbf{n}}

\def \bx{\mathbf{x}}
\def \bxh{\hat{\mathbf{x}}}
\def \bf{\mathbf{f}}
\def \bb{\mathbf{b}}
\def \bn{\mathbf{n}}

\def \M{\mathcal{M}}
\def \bU{\mathbf{U}}

\begin{document}

\title{Passive hydrodynamic synchronization of two-dimensional swimming cells}

\author{Gwynn J. Elfring}
\author{Eric Lauga\footnote{Corresponding author. Email: elauga@ucsd.edu}}
\affiliation{
Department of Mechanical and Aerospace Engineering, 
University of California San Diego,
9500 Gilman Drive, La Jolla CA 92093-0411, USA.}
\date{\today}
\begin{abstract}

Spermatozoa flagella are known to synchronize when swimming in close proximity. We use a model consisting of  two-dimensional sheets propagating transverse waves of displacement to demonstrate that fluid forces lead to such synchronization passively. Using two  distinct asymptotic descriptions (small amplitude and long wavelength), we derive the synchronizing dynamics analytically for arbitrarily shaped waveforms in Newtonian fluids, and show that phase locking will always occur for  sufficiently asymmetric shapes.
We characterize the effect of the geometry of the waveforms and the separation between the swimmers on the synchronizing dynamics, the final stable conformations, and  the energy dissipated by the cells. For two closely-swimming cells, synchronization always occurs at the in-phase or opposite-phase conformation, depending solely on the geometry of the cells. In contrast, the work done by the swimmers is always minimized at the  in-phase conformation. As the swimmers get further apart, additional fixed points arise at intermediate values of the relative phase. In addition, computations for large-amplitude waves using the boundary integral method reveal that the two asymptotic limits capture all the relevant physics of the problem. 
Our results provide a theoretical framework to address other hydrodynamic interactions phenomena  relevant to  populations of  self-propelled organisms.

\end{abstract}
\maketitle

\section{Introduction}

An often observed yet surprising physical phenomenon is the synchronization of the pendulums of grandfather clocks. When two such clocks are  located in close proximity,  forces transmitted through a medium connecting the two clocks can lead to their beating in perfect synchrony \cite{pikovsky02}. Similar synchronization can easily be obtained at home using two connected metronomes, with spectacular results \cite{metronome}. Still more fascinating is the many examples of synchrony which occur in the natural world, from pacemaker cells in a heart \cite{mirollo90}, to synchronously flashing fireflies \cite{buck38}.

\begin{figure}[b]
\includegraphics[width=0.4\textwidth]{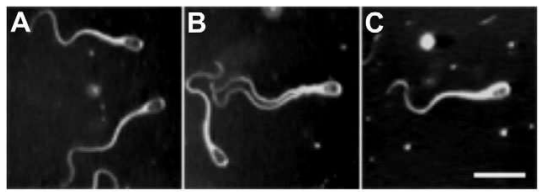}
\caption{(A to C) Time-sequence showing the synchronization of two swimming bull spermatozoa. Scale bar is $25\mu m$. Reproduced from Woolley et al. \cite{woolley09} with permission.}
\label{sync}
\end{figure}

One particularly interesting example of synchronization occurring in nature is the observed phase locking of the flagella of swimming eukaryotes such as spermatozoa \cite{hayashi98,riedel05,yang08,woolley09}. These cells, typically tens of microns long, actuate slender flagella beating periodically  in order to propel themselves in viscous fluids \cite{lighthill76,brennen77,lauga09b}. As illustrated in Fig.~\ref{sync} in the case of two bull spermatozoa, when two such cells swim in close proximity, their flagella are often observed to beat in synchrony -- so much so that in Fig.~\ref{sync}C the two flagella cannot even be distinguished \cite{woolley09}. This synchronization is biologically significant because it is observed to lead to an increased swimming speed for the co-moving cells, thereby providing a competitive advantage over cells which are not synchronized \cite{moore95,woolley09}. This behavior can arise purely passively, as is the case with the pendulums, but here the medium transmitting the forces is the fluid between the cells. While large systems of many bodies may be too complicated to address rigorously, and idealizations such as the Kuramoto model must be employed \cite{kuramoto75}, in this paper we consider the simple case of a pair of co-swimming two-dimensional cells. We show that the coupling between the bodies can be obtained directly by solving analytically and numerically the field equations for the surrounding flow and find the occurrence of passive hydrodynamic synchronization for all but the most symmetric flagellar waveforms.

G.I. Taylor first studied synchronizing flagellated cells by modeling  them as two dimensional sheets propagating sinusoidal waves of transverse displacement \cite{taylor51}. With this model, he found that, for a given swimming gait,  swimming in-phase synchronously is the conformation in which the cells swim while doing the least amount of work against the surrounding fluid. Left open was the question of whether the synchronization  would occur passively from a random initial phase shift between co-swimming cells. Subsequent numerical works using an immersed boundary method and multiparticle collision dynamics seem to indicate that indeed synchronization could occur due to hydrodynamic forces alone \cite{fauci90,fauci95,yang08}.

The phase locking of flagellated microorganisms is closely related to another important observed synchrony in nature, that of eukaryotic cilia. Cilia are short flagella typically lining the surface of a larger body and are found to beat in unison with a small constant phase difference giving rise to a collective motion described as metachronal waves \cite{brennen77}. This motion provides various biological functionality including fluid transport and locomotion \cite{blake75}. Several models with varying complexity have indicated that the synchronization which manifests as metachronal beating can occur due to fluid forces alone \cite{gueron97, lagomarsino03, lenz06, vilfan06} although, since individual cilia are not free-swimming but are attached to a substrate, synchronization can only occur with a load-dependent force generation. Similarly to cilia, there is an observed synchronization of the pairs of flagella used for propulsion on the alga \textit{Chlamydomonas} \cite{goldstein09}. Beyond  eukaryotic flagella and cilia, hydrodynamic interactions in bacterial flagella lead to the creation of flagella bundles propelling the cells forward as they swim, as well as the disruption of such bundles when the cells change their swimming direction \cite{mcnabb77, turner05, flores05}.

In this paper we return to the two dimensional model first proposed by Taylor (detailed in Sec.~\ref{sec:definition}), to describe the phase locking of swimming flagellated cells.  The simplicity of such a model allows one to address the problem analytically, to extract the relevant properties that such waves must possess in order to give rise to synchronization, and to determine precisely what states of dynamic equilibrium will occur. We first present geometrical arguments which show that Taylor's purely sinusoidal sheet cannot dynamically synchronize due to an excess of symmetry which, when coupled with the kinematic reversibility of the Stokes equations, prevents any relative motion between free-swimming cells (Sec~\ref{sec:symmetry}). Real flagella possess a front-back asymmetry and we show that this feature leads to the occurrence of synchronization. We accomplish this by allowing the sheets to pass completely general waveforms in our model. We then solve the problem analytically for two asymptotic limits, first when the amplitude of the waves is much smaller than their wavelength (Taylor's limit, Sec.~\ref{smallamp}), and then when the mean distance between the waves is much smaller than the wavelength (lubrication limit, Sec.~\ref{lublimit}). We also solve the problem numerically using the boundary integral formulation of the Stokes equations to demonstrate the validity of the analytic formulae and to address the synchronization of large-amplitude waveforms (Sec.~\ref{BIformulation}).

Our results show precisely how the geometry of the waveforms governs the synchronizing dynamics of the system  (Sec.~\ref{results}). We obtain simple formulae that dictate the time-evolution of the phase and the energy dissipation, and which indicate that while swimming in-phase results in a minimum of viscous dissipation it does not necessarily coincide with an equilibrium state, and indeed a dynamically stable state may maximize energy dissipation. In addition to the geometry of the waveforms, we demonstrate the importance of the separation of the sheets on the dynamics of the system. We show that the stable conformations (and the number of them) may change with the distance between the cells. Notably swimming cells with front-back asymmetry are shown synchronize into either a stable in-phase or opposite-phase conformation when in close proximity, while some cells when further apart are shown to synchronize with a fixed finite phase difference, reminiscent of ciliary phase locking. 
A discussion and summary of these results is offered in Sec.~\ref{conclusion}.

\section{Setup}\label{sec:definition}

\begin{figure}[t]
\includegraphics[width=0.6\textwidth]{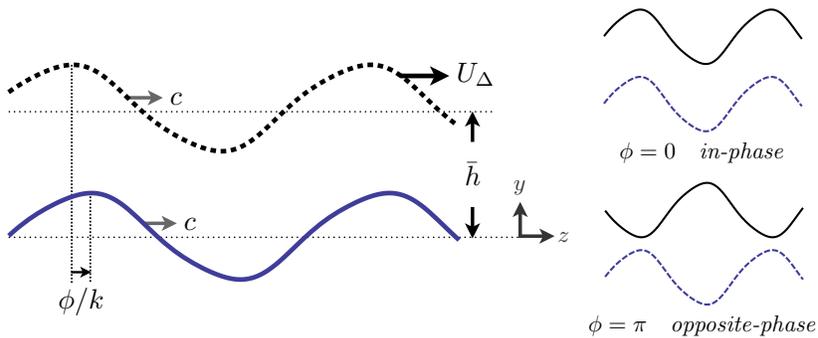}
\caption{System of parallel and identical two-dimensional infinite sheets in a frame moving with the lower sheet. The sheets are separated vertically by a mean distance $\hb$. The top sheet, behind the bottom sheet by a phase $\phi$ as measured along the $z$ axis,  moves to the right with a relative velocity $U_\Delta$.}
\label{system}
\end{figure}

Our system, as illustrated in Fig.~\ref{system}, consists of two parallel and identical infinite sheets, which we will call swimmers, separated by a mean distance $\hb$. The sheets both propagate waves of transverse displacement in the positive $z$ direction, with amplitude $a$ and speed $c=\omega/k$, where $\omega$ is the wave frequency and $k$ is the wavenumber, and have an initial phase difference $\phi_0=k\Delta z_0$ (denoted positive when the bottom sheet is shifted by $\phi_0$ along the positive $z$ direction with respect to the top sheet). By passing these waves the swimmers propel themselves in the $-z$ direction \cite{taylor51}. We consider the frame of reference moving with the bottom sheet, at speed $U$, and write the relative speed of the top sheet in the $z$ direction as $\UD$.

The instantaneous positions of the bottom ($y_1$) and top ($y_2$) sheets are thus given by
\begin{eqnarray}
y_1&=&a g\Big(k[z-ct]\Big),\\
y_2&=&\hb+a g\left(k\left[z-ct+\Delta z_0 -\int_{0}^{t}U_{\Delta}(t')dt'\right]\right),
\end{eqnarray}
where $g$ a function describing the arbitrary waveform of the swimmers, and $z$ is the axial coordinate in a frame moving with the lower sheet. We use the following dimensionless variables $z^*=zk$, $t^*=t\omega$, $u^*=u/c$, $v=v/\epsilon c$, with the ratio of the amplitude of the waves to their wavelength given by  $\epsilon=a k$. For convenience we use the wave variable $x^*=z^*-t^*$ and the instantaneous phase difference $\phi=\phi_0-k\int_{0}^{t^*} \UD^*(t')dt'$. Consequently the positions of the sheets in the moving frame are given simply by 
\begin{eqnarray}
y^*_1&=&\epsilon g(x^*),\\
y^*_2&=&\hb^*+\epsilon g(x^*+\phi),
\end{eqnarray}
where the arbitrary $2\pi$-periodic function $g$ can be written using Fourier series as
\begin{equation}
g(x^*) = \sum_{n=1}^\infty \alpha_n\cos(n x^*)+\sum_{n=1}^\infty \beta_n\sin(n x^*).
\end{equation}

Since we are concerned with the synchronization of microorganisms, we are in a low Reynolds number regime ($\text{Re}\sim10^{-4}$ for the bull spermatozoa in Fig.~\ref{sync}) where the fluid between the sheets is inertia free, and thus mechanical equilibrium for the stress tensor, $\bsigma^*$, is written as  $\nabla \cdot \bsigma^*={\boldsymbol{0}}$. Assuming an incompressible Newtonian flow we obtain the Stokes equations for the dimensionless velocity field, $\mathbf{u^*}=(u^*,v^*)$, and dynamic pressure, $p^*=p\epsilon^2/\mu\omega$, as
\begin{eqnarray}
\nabla^2\bu^*&=&\nabla p^*,\\
\nabla\cdot \bu^*&=&0.
\end{eqnarray}

Physically, if the sheets are not permitted to move relative to one another, i.e. if we set $\UD^*=0$, then there may arise a horizontal hydrodynamic force $f_x$ acting on the swimmers. Conversely, if we let the sheets move freely under the constraint that they are force free then there may be a nonzero evolution of the phase in time, given geometrically as $\UD^*=-d\phi/dt$. These two problems are of course related, as we will see, by the mobility $\mathcal{M}$, as $\UD=\mathcal{M}f_x$. In the case of a purely sinusoidal swimmer
(i.e. $\beta_1=1$, $\beta_n=0 \ n>1$ and $\alpha_n=0 \ \forall n$), 
G.I. Taylor \cite{taylor51} derived the swimming speed of a single sheet (the outer problem) and obtained 
\begin{eqnarray}
U^*=-\frac{1}{2}\epsilon^2\left(1-\frac{19}{16}\epsilon^2\right)+O(\epsilon^6).
\label{taylor}
\end{eqnarray}
In the rest of the paper we  drop the $^*$ notation for convenience.

\section{Symmetry}\label{sec:symmetry}

Before calculating the hydrodynamic forces between the swimmers, it is insightful to first consider the various symmetry properties of the problem, and their consequences on force generation and synchronization. 

\begin{figure}
\includegraphics[width=0.7\textwidth]{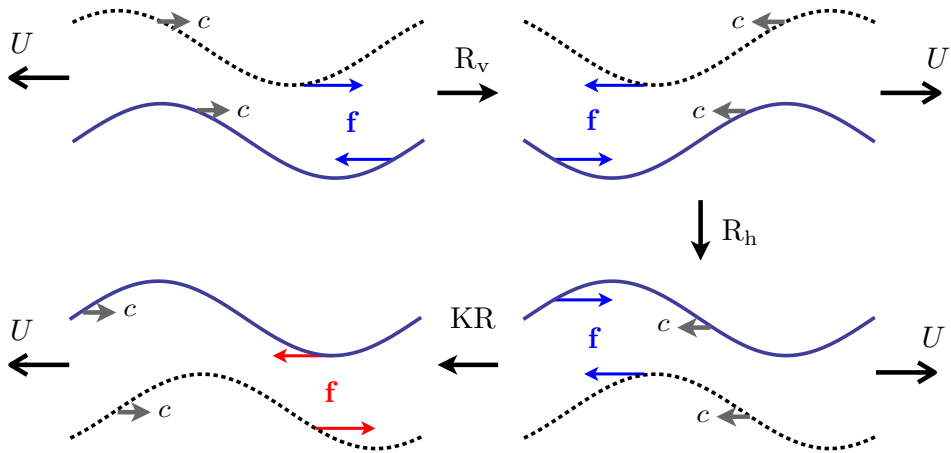}
\caption{A system of two identical and parallel swimmers which has a stabilizing force (top left) becomes destabilizing (bottom left), under two reflections -- first about the vertical axis ($\text{R}_{\text{v}}$) then about the horizontal axis ($\text{R}_{\text{h}}$) -- combined with an application of kinematic reversibility (KR), yet the boundary conditions remain identical, hence the force must be zero. Symmetric waveforms can thus not synchronize.
}
\label{symmetry}
\end{figure}

Suppose first that we have two swimmers, $g_1$ and $g_2$, whose shapes are such that $g_2$ is obtained from $g_1$ by a vertical axis reflection plus a horizontal axis reflection and a phase shift $\theta$ (which depends on the location of the vertical axis), i.e. $g_2(x)=-g_1(-x+\theta)$. In that case there can be no horizontal hydrodynamic force acting between the swimmers, and $f_x=0$. To prove this result, let us  assume that a force $\bf$ acts on the top sheet with $\UD=0$ (since $\nabla\cdot\bsigma=0$ the force on the bottom sheet must be equal and opposite in sign). We then perform a reflection of the entire conformation about the vertical axis then horizontal axis, followed by a reversal of the kinematics (see Fig.~\ref{symmetry} for an example). The resulting system is identical except the sign of the force has reversed, $\bf\rightarrow -\bf$, a contradiction unless $\bf=0$ (then $f_x=0$). In particular, if the sheets are identical, then there can be no synchronization if the identical shapes of the waveforms satisfy $g(x)=-g(-x+\theta)$. A subset of these shapes are sheets that are invariant under both vertical axis reflection $g(x)=g(-x+\theta)$ and horizontal axis reflection $g(x)=-g(x+\pi)$; the simplest example of such shape is a pure sinewave ($\beta_1=1$, $\beta_n=0 \ n>1$ and $\alpha_n=0 \ \forall n$), which is Taylor's original geometry \cite{taylor51,fauci90}. Since such an arrangement has both vertical and horizontal axis symmetry it will not passively synchronize in a Stokesian flow  \cite{elfring09}. Similar excessive geometrical symmetries have also been observed to curb any phase-locking in other swimmer models \cite{kim04,pooley07,putz09}.

A further generalization of the argument may be obtained by noting that in two dimensions the outer problem can balance no force and hence each side of the swimmer must be force free. This decoupling of the inner and outer problem means that it is only the fluid between the two sheets that drives synchronization, if any. Thus if  two swimmers do not phase-lock, a similar arrangement of more than two swimmers will not either -- a result that cannot be obtained by symmetry alone.

In order to possibly obtain a passive synchronization between the swimmers we must therefore either (1) have a geometry such that $g(x)\ne -g(-x+\theta)$, or (2) remove the kinematic reversibility of the flow equations. Since we are considering here microorganisms in a Newtonian fluid, the latter is a property of the problem that we cannot escape. If our model were at finite Reynolds number, or in a viscoelastic fluid, then this constraint would naturally be removed and symmetric swimmers could synchronize \cite{elfring10}. In a Stokesian flow we must thus have a geometrical asymmetry. 

Most swimming microorganisms, such as spermatozoa, possess a cell body  and thus have a very natural front-back asymmetry. In addition, some spermatozoa pass waves along their flagella which increase in amplitude from head to tail, leading to another type of front-back asymmetry \cite{rikmenspoel65}. In contrast, swimmers whose flagellar waveforms or body is asymmetric with respect to the horizontal axis experience viscous torques, and thus cannot swim straight. It is therefore natural for us to focus on waveforms  which are symmetric about the horizontal axis, but not the vertical.  As a result of this horizontal axis symmetry, the horizontal component of a force between the swimmers must be an odd function of the phase $\phi$, $f_x(-\phi)=-f_x(\phi)$, and thus there must always be a fixed point at $\phi=0$, i.e. $f(0)=0$. In addition,  because the force is 2$\pi$-periodic, then $\phi=\pi$ must also be another fixed point, i.e. $f(\pi)=0$.

As a side note, we observe that because of the kinematic reversibility of the Stokes equations, a change in the direction of wave propagation yields a reversal of forces $\mathbf{f}\rightarrow \mathbf{-f}$. Reversing the direction of wave propagation is geometrically equivalent to reversing the front-back asymmetry of the waveforms which must therefore also reverse the forces on the swimmer.

In order to gain physical intuition in the synchronization process, we now characterize the force generation and subsequent synchronization between the two sheets analytically by focusing on two  asymptotic limits. We first consider in Sec.~\ref{smallamp}  the limit in which the amplitude of the traveling waves is much smaller than their wavelength. The limit in which the distance between the swimmers is much smaller than their wavelength will then be considered in Sec.~\ref{lublimit}.  Additionally we solve the problem numerically using the boundary integral formulation of the Stokes equations in Sec.~\ref{BIformulation} to validate our asymptotics and address large-amplitude swimming.

\section{Small Amplitude Expansion}\label{smallamp}

Because the model is two dimensional we may introduce the stream function formulation and write $\bu=\{\partial\psi/\partial y, -\partial\psi/\partial x\}$. In this manner the continuity equation is automatically satisfied and the Stokes equations reduce to a biharmonic equation in the stream function
\begin{eqnarray}
\nabla^4\psi=0.
\label{biharmonic}
\end{eqnarray}

We assume in this section  that the amplitude of the traveling wave is much smaller than their wavelength, $\epsilon \ll 1$, and look to solve this problem by seeking a regular perturbation expansion in powers of $\epsilon$, $\psi=\sum\epsilon^n\psi_n$. Because of the symmetry of the problem there is no difference in the boundary conditions if we change $\epsilon\rightarrow -\epsilon$ as this is equivalent to taking $x\rightarrow x+\pi$, this then naturally precludes the possibility of a synchronizing force appearing at all odd powers in $\epsilon$.

\subsection{Boundary Conditions}
We wish to prescribe a wave of transverse displacement to each sheet. However, doing  so requires the material composing the sheets to be extensible as material points will accelerate relative one another. If we wish to forbid this relative motion, we may require the sheet to pass waves in an inextensible fashion. This may be visualized as material points moving along a conveyor of static shape, when observing the sheet in the wave frame \cite{lighthill75,childress81}.

For extensible sheets the boundary conditions are given simply by the time derivatives of the waveforms namely
\begin{subeqnarray}\label{ext}
u\mid_{y= y_1} &=& 0,\\
v\mid_{y=y_1} &=& \partial y_1/\partial t,\\
u\mid_{y= y_2} &=& 0,\\
v\mid_{y=y_2} &=& \partial y_2/\partial t.
\end{subeqnarray}

For inextensible inextensible sheets the boundary conditions are given by
\begin{subeqnarray}
u\mid_{y= y_1} &=& 1-\SR\cos\theta\mid_{y= y_1},\\
v\mid_{y= y_1} &=& -\SR\sin\theta\mid_{y= y_1},\\
u\mid_{y= y_2} &=& 1-\SR\cos\theta\mid_{y= y_2}+\UD,\\
v\mid_{y= y_2} &=& -\SR\sin\theta\mid_{y= y_2}.
\end{subeqnarray}
where the angle, $\theta$, is defined by $\tan\theta = \partial y/\partial x$ hence
\begin{eqnarray}
\cos\theta &=& \frac{1}{\sqrt{1+(y')^2}},\\
\sin\theta &=& y'\cos\theta,
\end{eqnarray}
and the material velocity (in the wave frame), $\SR$, is ratio of the length of the sheet to its wavelength multiplied by the wave speed, or
\begin{equation}
\SR=\frac{1}{2\pi}\int_{0}^{2\pi}{\sqrt{1+\left(\frac{\partial y}{\partial x}\right)^2}dx}.
\end{equation}

\subsection{Expansion}
Since we know that an expansion can yield a synchronizing force only at even powers in amplitude one would hope to see a relative force generated at order $\epsilon^2$. We actually show below that for any waveform $g(x)$, the force is zero at order $\epsilon^2$, and hence a perturbation expansion must carried out to order $\epsilon^4$ in order to obtain the synchronizing dynamics.

\subsubsection{Flow at $O(\epsilon)$}

The governing equation at $O(\epsilon)$ is
\begin{eqnarray}
\nabla^4\psi_1=0,
\end{eqnarray}
and the boundary conditions are given by
\begin{subeqnarray}\label{bcexp1}
{\nabla}\psi_1\mid_{y= 0} &=& {\nabla}g(x),\\
{\nabla}\psi_1\mid_{y= \hb} &=& {\nabla}g(x+\phi)+\be_y \UDone,
\end{subeqnarray}
where $\be_y$ denotes the unit vector in the $y$ direction. We note that the boundary conditions at $O(\epsilon)$ are the same for both extensible and inextensible motion.

The biharmonic equation can be solved by repeated separation of variables. The general solution may be expressed as
\begin{align}
\psi_1(x,y) =& A_{1,0}+B_{1,0}y+C_{1,0}y^2+D_{1,0}y^3+ (E_{1,0}+F_{1,0}y+G_{1,0}y^2+H_{1,0}y^3)x \nonumber \\
&+ \sum_{n=1}^\infty \Big[(A_{1,n}+B_{1,n} y)\sinh(n y)+(C_{1,n}+D_{1,n} y)\cosh(n y)\Big]\cos(n x) \nonumber \\
&+\sum_{n=1}^\infty \Big[(E_{1,n}+F_{1,n} y)\sinh(n y)+(G_{1,n}+H_{1,n} y)\cosh(n y)\Big]\sin(n x),
\label{generalsolution}
\end{align}
where for the constants $A$ through $H$, the first subscript refers to the order in the expansion (here, first order) and the second refers to the corresponding Fourier mode. We can immediately discard the terms linear in $x$ due to the periodicity of the problem.

From the first order boundary conditions, Eq.~\eqref{bcexp1}, we get that the solution to the biharmonic equation may be written analytically as
\begin{eqnarray}
\psi_1=a_{1,0}(y)+\sum_{n=1}^{\infty}\Big[a_{1,n}(y)\cos(nx)+b_{1,n}(y)\sin(nx)\Big],
\end{eqnarray}
where
\begin{eqnarray}
a_{1,0}(y) &=& \frac{\left(\UDone-3 \text{D}_{1,0} \hb^2\right) y^2}{2 h}+\UDone y^3,\\
a_{1,n}(y) &=& 2P_{n}(y)\big(\alpha_n\cos(n\phi)+\beta_n\sin(n\phi)\big)+\alpha_n Q_{n}(y),\\
b_{1,n}(y) &=& 2P_{n}(y)\big(\beta_n\cos(n\phi)-\alpha_n\sin(n\phi)\big)+\beta_n Q_{n}(y),
\end{eqnarray}
and
\begin{eqnarray}
P_{n}(y) &=& \left[\frac{ n^2\hb y \cosh(n\hb)+\sinh(n\hb)n y }{2 n^2\hb^2-2\sinh^2(n\hb)}\right]\cosh(ny)\nonumber \\
&&-\left[\frac{\left(1+\hb n^2 y\right) \sinh(n\hb)+h n \cosh(n\hb)}{2 n^2\hb^2-2\sinh^2(n\hb)}\right]\sinh(ny),\slabel{Pn}\\
Q_{n}(y) &=& \left[\frac{2n\hb+2n y \sinh^2(n\hb)+\sinh(2n\hb)}{2 n^2\hb^2-2\sinh^2(n\hb)}\right]\sinh(ny)\nonumber \\
&&+\left[1-\frac{2\hb n^2y+n y \sinh(2 n\hb) }{2 n^2\hb^2-2\sinh^2(n\hb)}\right]\cosh(ny)\slabel{Qn}.
\end{eqnarray}

In order to solve for the unknown constant $D_{1,0}$ we turn to dynamical considerations. To compute the force on the bottom sheet we note that we are free to move the  integral along the surface of the sheet $S$, to any surface parallel to the $x$ axis. This  can be shown by integrating $\nabla\cdot\bsigma=0$ over the area between the sheet and any such surface and using the  periodicity of the problem.
 Alternatively, this can be shown by expanding as follows
\begin{subeqnarray}\label{24}
f_x\mid_{y=y_1}&=&\be_x\cdot\int_{S}\left(\bsigma\cdot\bn\right)\mid_{y=y_1} dS \\
&=&\int_{0}^{2\pi}\Big(\sigma_{xy}-\epsilon g'(x)\sigma_{xx}\Big)\mid_{y=y_1} dx\\
&=&\int_{0}^{2\pi}\left(\sigma_{xy}-\sum_{n=1}^{\infty}\frac{\partial}{\partial x}\left[\epsilon^{n-1} g(x)^{n-1}\frac{\partial^{n-1}\sigma_{xx}}{\partial y^{n-1}}\right]\right)\mid_{y=0} dx\\
&=&\int_{0}^{2\pi}\sigma_{xy}\mid_{y=0} dx.
\end{subeqnarray}
We will use the result given by Eq.~\eqref{24} repeatedly throughout the paper.

Using the above we find that the force on the bottom sheet to $O(\epsilon)$ is 
\begin{subeqnarray}
f_{1x}\mid_{y=y_1} &=& \int_0^{2\pi}\left(\frac{\partial^2\psi_1}{\partial y^2}-\frac{\partial^2\psi_1}{\partial x^2}\right) \mid_{y=0}dx\\
  &=& 2\pi a_{1,0}''(0)\\
  &=& -6 D_{10} \hb \pi +\frac{2 \pi}{\hb}\UDone,
\end{subeqnarray}
while the force on the upper sheet is similarly
\begin{subeqnarray}
f_{1x}\mid_{y=y_2}  &=& -2\pi a_{1,0}''(\hb)\\
  &=& -6 D_{10} \hb \pi -\frac{2 \pi}{\hb}\UDone.
\end{subeqnarray}
Hence we see that only the second derivative of the zeroth Fourier mode contributes to the force.

Finally, integrating mechanical equilibrium, ${\nabla}\cdot\boldsymbol{\sigma}=0$,  between the two sheets
leads to the equality  
\begin{eqnarray}
\mathbf{f}\mid_{y=y_1}+\mathbf{f}\mid_{y=y_2}=0
\end{eqnarray}
where $\mathbf{f}=\int_S \bsigma\cdot\bn dS$. Taking the $x$-component we find $f_{1x}\mid_{y=y_1}=-f_{1x}\mid_{y=y_2}$ and in order to satisfy this relationship we must have $D_{1,0}=0$. The force on the \textit{top} sheet is therefore
\begin{eqnarray}
f_{1x}=-\frac{2\pi}{\hb}\UDone.
\label{forcebalance}
\end{eqnarray}
If $\UDone=0$ then there is no phase locking force. Conversely if the sheets are force free then $\UDone=0$. There is thus no synchronization at $O(\epsilon)$, as  expected from the $\epsilon\rightarrow -\epsilon$ symmetry. 

\subsubsection{Flow at $O(\epsilon^2)$}
The governing equation at $O(\epsilon^2)$ is
\begin{eqnarray}
\nabla^4\psi_2=0,
\end{eqnarray}
while the boundary conditions are given by
\begin{subeqnarray}
{\nabla}\psi_2\mid_{y= 0} &=& -g(x) {\nabla}\left(\frac{\partial \psi_1}{\partial y}\right)\mid_{y= 0}\nonumber\\
&& +\frac{\be_y}{2}\left[g'(x)^2-\frac{1}{2\pi}\int_{0}^{2\pi}g'(x)^2dx\right],\\
{\nabla}\psi_2\mid_{y= \hb} &=& \be_y\UDtwo-g(x+\phi) {\nabla}\left(\frac{\partial \psi_1}{\partial y}\right)\mid_{y= \hb}\nonumber\\ 
&&+\frac{\be_y}{2}\left[g'(x+\phi)^2-\frac{1}{2\pi}\int_{0}^{2\pi}g'(x+\phi)^2dx\right],\quad \quad 
\end{subeqnarray}
where the terms in the square brackets represent the contribution from the inextensibility constraint. 

We may write the solution to the biharmonic equation as
\begin{equation}
\psi_2=a_{2,0}(y)+\sum_{n=1}^\infty a_{2,n}(y)\cos(n x)+\sum_{n=1}^\infty b_{2,n}(y)\sin(n x).
\end{equation}
The zeroth mode is given by
\begin{equation}
a_{2,0}(y)=\text{D}_{2,0} y^3+\frac{y^2 \left(\UDtwo+u_{2h}-u_{20}-3 \text{D}_{2,0} \hb^2\right)}{2\hb}+y u_{20},
\end{equation}
where we define for $O(\epsilon^m)$ in the expansion  
\begin{subeqnarray}\label{um}
u_{m0}&=&\frac{1}{2\pi}\int_{0}^{2\pi}\frac{\partial\psi_m}{\partial y}\mid_{y= 0}dx,\\
u_{mh}+\UDm&=&\frac{1}{2\pi}\int_{0}^{2\pi}\frac{\partial \psi_m}{\partial y}\mid_{y= \hb}dx,
\end{subeqnarray}
as the mean components of the horizontal boundary conditions. The vertical boundary conditions cannot have a mean component and therefore do not contribute to the zeroth Fourier mode of the solution.

Again to resolve the unknown constant we turn to dynamical considerations. Utilizing Eq.~\eqref{forcebalance} and evaluating at order $\epsilon^2$ 
we obtain 
\begin{eqnarray}
a_{20}''(0)=a_{20}''(\hb),
\end{eqnarray}
therefore $D_{2,0}=0$. The force on the top sheet is then given as
\begin{equation}
f_{2x}=\frac{2\pi}{\hb}(u_{20}-u_{2h}-\UDtwo).
\end{equation}
The mean components of the horizontal boundary conditions must then be evaluated, the lower
\begin{eqnarray}
u_{20}=\frac{1}{2\pi}\int_{0}^{2\pi}\left(-g(x)\left(\frac{\partial^2 \psi_1}{\partial y^2}\right)\mid_{y= 0}+\frac{1}{2}\left[g'(x)^2-\frac{1}{2\pi}\int_{0}^{2\pi}g'(t)^2dt\right]\right) dx.
\end{eqnarray}
The term in the square brackets clearly integrates to zero hence we are left with
\begin{eqnarray}
u_{20}=-\frac{1}{2\pi}\int_{0}^{2\pi}g(x)\left(\frac{\partial^2 \psi_1}{\partial y^2}\right)\mid_{y= 0}dx,
\end{eqnarray}
which, using orthogonality of Fourier modes,  gives
\begin{eqnarray}
u_{20}=-\frac{1}{2}\sum_{n=1}^{\infty}\Big[\alpha_n a_{1,n}''(0)+\beta_n b_{1,n}''(0)\Big].
\end{eqnarray}
Similarly $u_{2h}$ is given by
\begin{eqnarray}
u_{2h}=-\frac{1}{2\pi}\int_{0}^{2\pi}g(x+\phi)\left(\frac{\partial^2 \psi_1}{\partial y^2}\right)\mid_{y= \hb}dx, 
\end{eqnarray}
which may be evaluated to give
\begin{eqnarray}
&\displaystyle u_{2h}=-\frac{1}{2}\sum_{n=1}^{\infty}\Big[&(\alpha_n a_{1,n}''(\hb)+\beta_n b_{1,n}''(\hb))\cos(n\phi)\nonumber\\
&& +(\beta_n a_{1,n}''(\hb)-\alpha_n b_{1,n}''(\hb))\sin(n\phi)\,\Big].
\end{eqnarray}
Further, by considering $P_{n}$ and $Q_{n}$, given by Eq.~\eqref{Pn} and Eq.~\eqref{Qn} respectively, 
and observing that $2P_{n}''(\hb)=-Q_{n}''(0)$ and $2P_{n}''(0)=Q_{n}''(h)$
it can be shown that
\begin{eqnarray}
&&-\Big[\alpha_n a_{1,n}''(0)+\beta_n b_{1,n}''(0)\Big]+\Big[\alpha_n a_{1,n}''(\hb)+\beta_n b_{1,n}''(\hb)\Big]\cos(n\phi) \nonumber\\ 
&&+\Big[\beta_n a_{1,n}''(\hb)-\alpha_n b_{1,n}''(\hb)\Big]\sin(n\phi)\nonumber \\
&=&(\alpha_n^2+\beta_n^2)\left[2P_{n}''(\hb)-Q_{n}''(0)-\cos(n\phi)\Big(2P_{n}''(0)-Q_{n}''(\hb)\Big)\right]=0, 
\end{eqnarray}
for all $n$. 

The force on the top sheet is then equal to
\begin{eqnarray}
f_{2x}=-\frac{2\pi}{\hb}\UDtwo.
\end{eqnarray}
Here again we see that when we allow the swimmers to move in a force free manner then $\UDtwo=0$ and hence there is no synchronization at $O(\epsilon^2)$. Note that we have not specified the Fourier coefficients of the of the waveform, and this result is therefore valid for any waveform  $g(x)$.

Sadly, due to the $\epsilon \rightarrow -\epsilon$ symmetry of the model there cannot be any force at $O(\epsilon^3)$, and therefore we expect the force to arise at best at $O(\epsilon^4)$.

\subsubsection{Flow at $O(\epsilon^3)$}
The third order component of Eq.~\eqref{biharmonic} is
\begin{equation}
\nabla^4\psi_3=0.
\label{biharmonic3}
\end{equation}

With the third order boundary conditions
\begin{subeqnarray}
\nabla\psi_3\mid_{y= 0} \ &=& -g(x)\left[\nabla\left(\frac{\partial \psi_2}{\partial y}\right)+\frac{g(x)}{2}\nabla\left(\frac{\partial^2 \psi_1}{\partial y^2}\right)\right]\mid_{y=0} \nonumber\\
&& -\be_x\frac{g'(x)}{2}\left[g'(x)^2-\frac{1}{2\pi}\int_{0}^{2\pi}g'(x)^2dx\right],\\
\nabla\psi_3\mid_{y= \hb} \ &=& \be_y\UDthree-g(x+\phi)\left[\nabla\left(\frac{\partial \psi_2}{\partial y}\right)+\frac{g(x+\phi)}{2}\nabla\left(\frac{\partial^2 \psi_1}{\partial y^2}\right)\right]\mid_{y=\hb}\nonumber \\
&&-\be_x\frac{g'(x+\phi)}{2}\left[g'(x+\phi)^2-\frac{1}{2\pi}\int_{0}^{2\pi}g'(x+\phi)^2dx\right].
\end{subeqnarray}
The solution can shown to be of the form
\begin{eqnarray}
\psi_3=a_{3,0}(y)+\sum_{n=1}^{\infty}\Big[a_{3,n}(y)\cos(nx)+b_{3,n}(y)\sin(nx)\Big],
\end{eqnarray}
where
\begin{eqnarray}
a_{3,0}(y) &=& \text{D}_{3,0} y^3+\frac{\left(\UDthree+u_{3h}-u_{30}-3 \text{D}_{3,0} \hb^2\right) y^2}{2 \hb}+yu_{30}.
\end{eqnarray}

Again we must resort to dynamical considerations to solve for $\text{D}_{3,0}$ and $\UDthree$. From Eq.~\eqref{forcebalance}
we find $a_{3,0}''(0) = a_{3,0}''(\hb)$ and hence $\text{D}_{3,0}=0$. The force again takes the form
\begin{eqnarray}
f_{3x}=\frac{2\pi}{\hb}\left(u_{30}-u_{3h}-\UDthree\right).
\end{eqnarray}
If the swimmers are force free we see $\UDthree=u_{30}-u_{3h}$ but due to the $\epsilon\rightarrow-\epsilon$ symmetry of the geometry we must have $\UDthree=0$ (in the example we consider in Sec.~\ref{results} $u_{30}=u_{3h}=0$).

\subsubsection{Flow at $O(\epsilon^4)$}
The fourth order component of Eq.~\eqref{biharmonic} is
\begin{equation}
\nabla^4\psi_4=0.
\label{biharmonic4}
\end{equation}

The boundary conditions at fourth order are given by
\begin{subeqnarray}
\nabla\psi_4\mid_{y= 0} \ &=& -g(x)\left[\nabla\left(\frac{\partial \psi_3}{\partial y}\right)+\frac{g(x)}{2}\nabla\left(\frac{\partial^2 \psi_2}{\partial y^2}\right)+\frac{g(x)^2}{6}\nabla\left(\frac{\partial^3 \psi_1}{\partial y^3}\right)\right]\mid_{y=0}\nonumber \\
&&-\be_y\frac{g'(x)^2}{4}\left[\frac{3}{2}g'(x)^2-\frac{1}{2\pi}\int_{0}^{2\pi}g'(x)^2dx\right]+\frac{\be_y}{16\pi}\int_{0}^{2\pi}g'(x)^4 dx,\\
\nabla\psi_4\mid_{y=\hb} \ &=& -g(x+\phi)\left[\nabla\left(\frac{\partial \psi_3}{\partial y}\right)+\frac{g(x+\phi)}{2}\nabla\left(\frac{\partial^2 \psi_2}{\partial y^2}\right)+\frac{g(x+\phi)^2}{6}\nabla\left(\frac{\partial^3 \psi_1}{\partial y^3}\right)\right]\mid_{y=\hb}\nonumber \\
&&-\be_y\frac{g'(x+\phi)^2}{4}\left[\frac{3}{2}g'(x+\phi)^2-\frac{1}{2\pi}\int_{0}^{2\pi}g'(x+\phi)^2dx\right]\nonumber \\
&& +\be_y\UDfour+\frac{\be_y}{16\pi}\int_{0}^{2\pi}g'(x+\phi)^4 dx.\quad 
\end{subeqnarray}

The solution is of the form
\begin{eqnarray}
\psi_4=a_{4,0}(y)+\sum_{n=1}^{\infty}\Big[a_{4,n}(y)\cos(nx)+b_{4,n}(y)\sin(nx)\Big].
\end{eqnarray}
However we are only interested in its mean component, i.e.
\begin{equation}
a_{40}(y)=\text{D}_{4,0} y^3+\frac{y^2 \left(\UDfour-3 \text{D}_{4,0} \hb^2+u_{4h}-u_{40}\right)}{2\hb}+y u_{40},
\end{equation}
From Eq.~\eqref{forcebalance} we obtain $a_{4,0}''(0)=a_{4,0}''(\hb)$ and hence $\text{D}_{4,0}=0$. The force on the upper sheet is then
\begin{equation}
f_{4x}=\frac{2 \pi}{\hb}(u_{40}-u_{4h}-\UDfour).
\end{equation}
Setting $\UD=0$ gives rise to a phase-locking force in the static case, $f_x^s$ (we use the superscript $s$ to avoid confusion), given by
\begin{eqnarray}\label{f4}
f_x^s=\epsilon^4\frac{2\pi}{\hb}(u_{40}-u_{4h})+O(\epsilon^6).
\end{eqnarray}
For  free-swimming we thus see that the relative swimming sped is given by
\begin{eqnarray}
\UD=\frac{\hb}{2\pi}f_x^s.
\end{eqnarray}
Noting that $d\phi/dt=-\UD$ we therefore get an equation for the time-evolution of the phase as
\begin{eqnarray}
\frac{d\phi}{dt}=-\frac{\hb}{2\pi}f_x^s=\epsilon^4(u_{4h}-u_{40})+O(\epsilon^6).
\label{dphiexp}
\end{eqnarray}
Importantly, the formulae for $u_{40}$ and $u_{4h}$, defined in Eq.~\eqref{um}, are too unwieldy for the most enterprising appendix even for simple $g(x)$, and hence are not stated explicitly (although straightforward to obtain with a symbolic calculation package).  In Sec.~\ref{results} these analytical results for both the phase locking force and the dynamical problem will be compared with a full numerical solution using the boundary integral formulation.

\subsection{Energy dissipation}
\label{smallenergy}
The energy dissipation rate between two sinusoidal sheets was originally computed by Taylor at leading order in the wave amplitude \cite{taylor51}. Here we restate his results for a general traveling wave. The energy dissipation per unit width in the fluid is equal to the rate of work of the sheets against the fluid
\begin{eqnarray}
\dot{E}=-\int_S (\bu\cdot\bsigma\cdot\bn)\mid_{y=y_1}dS-\int_S (\bu\cdot\bsigma\cdot\bn)\mid_{y=y_2}dS.
\end{eqnarray}
Expanding the integral in $\epsilon$ we find to leading order
\begin{eqnarray}
\dot{E}&=&\epsilon^2\int_{0}^{2\pi}g'(x)\left(-p_1+2\frac{\partial v_1}{\partial y}\right)\mid_{y=0}dx\nonumber\\
&& -\epsilon^2\int_{0}^{2\pi}g'(x+\phi)\left(-p_1+2\frac{\partial v_1}{\partial y}\right)\mid_{y=\hb}dx.
\end{eqnarray}
Expressing the pressure in terms of the stream function and integrating by parts yields
\begin{eqnarray}
\dot{E}=-\epsilon^2\int_{0}^{2\pi}g(x)\frac{\partial^3\psi_1}{\partial y^3}\mid_{y=0}dx+\epsilon^2\int_{0}^{2\pi}g(x+\phi)\frac{\partial^3\psi_1}{\partial y^3}\mid_{y=\hb}dx.
\end{eqnarray}
We already know the form of these integrals (indeed they are equal) from the analysis of the force at $O(\epsilon^2)$, and we find
\begin{eqnarray}
\dot{E}&=&\pi\epsilon^2\sum_{n=1}^{\infty}(\alpha_n^2+\beta_n^2) \nonumber \\
&&\times\Big[Q_n'''(0)-2P_n'''(h) -\cos(n\phi)\left(Q_n'''(h)-2P_n'''(0)\right) \Big],
\end{eqnarray}
which we can evaluate to get
\begin{eqnarray}
\dot{E}=2\pi\epsilon^2\sum_{n=1}^{\infty}n^3(\alpha_n^2+\beta_n^2)\left[A(n\hb)-\cos(n\phi)B(n\hb)\right],
\label{energyexp}
\end{eqnarray}
where
\begin{subeqnarray}
A(\xi)&=&\frac{2\xi+\sinh2\xi}{\sinh^2\xi-\xi^2},\\
B(\xi)&=&\frac{2\xi\cosh\xi+2\sinh\xi}{\sinh^2\xi-\xi^2}.
\label{AB}
\end{subeqnarray}
Setting $\beta_1=1$ and all other coefficients to zero in the above yields Taylor's result for pure sinewaves \cite{taylor51}.

In the limit $\xi\rightarrow\infty$, we see that $A\rightarrow2$, $B\rightarrow 0$, and the ratio $B/A$ decays exponentially. This tells us what we intuitively expect: When $\hb$ is large, the phase difference has little effect on the energy dissipation, and also  the phase difference has a weaker effect on the energy dissipated by higher Fourier modes. Conversely,  we expect that when the separation is small, the phase angle would have an large influence on the rate of working of the swimmers, and indeed when $\xi\rightarrow 0$, we zee $B/A\rightarrow 1$ as both $A,B\rightarrow 12\xi^{-3}$ (keeping in mind that we have implicitly assumed $\epsilon\ll\hb$). 

Importantly, because $A$ and $B$ are both positive and monotonically decaying functions with $\xi$, we know that  in-phase swimming, $\phi=0$, is a global minimum for the energy dissipated in the fluid. In addition, given that we have the symmetry $g(x+\pi)=-g(x)$, this restricts us to odd Fourier modes, and thus the out-of-phase configuration, $\phi=\pi$, is a global maximum. Taylor's dissipation argument \cite{taylor51} extends thus to arbitrary waveforms.

\section{Lubrication Limit}
\label{lublimit}

A second insightful limit to consider is the one in which  the sheets are so close together that their mean separation is much smaller than the wavelength of the oscillations, $\bar{h}\ll \lambda$. In this lubrication limit the Stokes equations are substantially simplified, permitting  analytical solutions. The main results of this section were previously summarized in a letter by the authors \cite{elfring09}.

\subsection{Lubrication equations}
In order to facilitate this limit we must rescale the governing equations. We nondimensionalize vertical distances by $y^*=y/\bar{h}$, and horizontal distances $z^*=kz$, while assuming that $\delta =k\bar{h}\ll 1$. The instantaneous position of the sheets is therefore given by $y^*_1=a^*g(x^*)$ and $y^*_2=1+a^*g(x^*+\phi)$, where $a^*=a/h$ and again $x^*=z^*-t^*$ is the wave variable. Nondimensionalizing the continuity equation we find that if the horizontal velocity is given by $u=cu^*$ then the vertical velocity must be $v=\delta c v^*$. The Stokes equations then yield the lubrication equations to leading order in $\delta$:
\begin{eqnarray}
\frac{\partial u^*}{\partial x^*}&=&-\frac{\partial v^*}{\partial y^*},\\
\frac{\partial p^*}{\partial y^*}&=&0,\\
\frac{dp^*}{dx^*}&=&\frac{\partial^2 u^*}{\partial y^{*2}},
\end{eqnarray}
where $p^*=\delta^2p/ \mu\omega$. Forces (per unit depth) are nondimensionalized as $f^*= f\delta/\mu c$, while energy dissipation rate per unit depth is $\dot{E}^*=\delta^2\dot{E}/\mu\omega c\hb$. We note that if $g$ approaches a singular geometry we would leave the realm of validity of the lubrication approximation \cite{wilkening08,wilkening09}. We now drop the $^*$ notation for convenience.

We look to solve this problem in a frame moving with the wave speed of the bottom sheet. The boundary conditions in the lubrication limit are then given by
\begin{subeqnarray}
u(x,y=y_1)=-1,\\
v(x,y=y_1)=-y_1',\\
u(x,y=y_2)=U_\Delta-1,\\
v(x,y=y_2)=-y_2'.
\end{subeqnarray}
Hence we see that in the lubrication limit the boundary conditions are identical to those of an extensible sheet. The full problem, regardless of whether extensible or inextensible boundary conditions are used, will collapse to the following lubrication results in the limit $\delta\ll 1$.

Given the above boundary conditions, the solution for the velocity field is found to be
\begin{eqnarray}
u(x,y)=\frac{1}{2}\frac{dp}{dx}(y-y_1)(y-y_2)+\UD\frac{y-y_1}{y_2-y_1}-1.
\label{u}
\end{eqnarray}
If one integrates the continuity equation one finds
\begin{eqnarray}
\int_{y_1}^{y_2}\frac{\partial u}{\partial x} dy  =y_2'-y_1',
\end{eqnarray}
which then gives
\begin{eqnarray}
\frac{dQ}{dx}=\UD\frac{dy_2}{dx}\cdot
\label{flowrate}
\end{eqnarray}
If no relative motion, $\UD=0$, then the flow rate $Q$ between the sheets is constant.

\subsection{Hydrodynamic force}

We first characterize  the force generated when $\UD=0$ in order to determine the location and nature of the  fixed points for the phase difference between the swimmers. With $\UD=0$ then $Q=\text{const.}$, and  we find
\begin{eqnarray}
Q=\int_{y_1}^{y_2}u dy=-\frac{1}{12}\frac{dp}{dx}h^3-h,
\end{eqnarray}
where $h=y_2-y_1$. We now exploit the periodicity of the system to obtain the value of $Q$ by noting that
\begin{eqnarray}
\int_{0}^{2\pi}\frac{dp}{dx}dx=-12\int_{0}^{2\pi}\frac{1}{h^2}dx-12Q\int_{0}^{2\pi}\frac{1}{h^3}dx=0.
\end{eqnarray}
We thus have
\begin{eqnarray}
Q=-\frac{I_2}{I_3},
\end{eqnarray}
where
\begin{eqnarray}
I_j&=&\int_{0}^{2\pi}h^{-j}dx.
\end{eqnarray}
The pressure gradient is therefore given by
\begin{eqnarray}
\frac{dp}{dx}=12\left(\frac{I_2}{h^3I_3}-\frac{1}{h^2}\right)\cdot
\label{pressure}
\end{eqnarray}

The force per unit depth on the upper sheet is given by
\begin{eqnarray}\label{fsx_lub}
f_x=\be_x\cdot\int_{S}\bsigma\cdot\bn\mid_{y=y_2} dS,
\end{eqnarray}
where the curve S is defined by $y=y_2$. Evaluating Eq.~\eqref{fsx_lub}  gives
\begin{eqnarray}
f_x&=&\int_{0}^{2\pi}\left[\frac{dy_2}{dx}\left(-p+2\delta^2\frac{\partial u}{\partial x}\right)-\left(\frac{\partial u}{\partial y}+\delta^2\frac{\partial v}{\partial x}\right)\right]\mid_{y=y_2}dx.
\end{eqnarray}
We keep only the $O(1)$ terms in the lubrication limit $\delta \ll 1$ which yields
\begin{eqnarray}
f_x&=&-\int_{0}^{2\pi}\left(\frac{dy_2}{dx}p+\frac{\partial u}{\partial y}\right)\mid_{y=y_2}dx.
\end{eqnarray}
Exploiting the periodicity of the problem through integration by parts \cite{chan05} allows us to recast the force as
\begin{eqnarray}\label{fsx_lub_final}
f_x=\int_{0}^{2\pi}\left(y_2\frac{dp}{dx}-\frac{\partial u}{\partial y}\right)\mid_{y=y_2}dx.
\end{eqnarray}
Substituting in Eq.~\eqref{u} and Eq.~\eqref{pressure}, and noting any constant multiplying the pressure gradient may be discarded, we find the force to be given by 
\begin{eqnarray}
f_x=6a\int_{0}^{2\pi}\left(\frac{I_2}{h^3I_3}-\frac{1}{h^2}\right)\big[g(x+\phi)+g(x)\big]dx.
\label{lubforce}
\end{eqnarray}

\subsection{Fixed points}

By symmetry, we found earlier that  there are always fixed points at $\phi=0,\pi$. This is easily confirmed by  evaluating Eq.~\eqref{lubforce}.  For $\phi=0$, $h$ is constant, and thus ${I_2}/{h^3I_3}-{1}/{h^2}=0$, leading to $f_x=0$;  for $\phi=\pi$, we have $g(x+\pi)+g(x)=0 $ by symmetry, and again $f_x=0$.

In order to determine their stability, we can expand the force, Eq.~\eqref{lubforce},  about these fixed points. Letting $\phi=\phi_0+\phi'$ where $\phi'\ll 1$ we obtain near the in-phase fixed point, $\phi_0=0$,
\begin{eqnarray}
f_{x_0}=-72a^4\phi'^3\int_{0}^{2\pi}g(x)g'(x)^3dx+O(\phi'^4).
\label{force0lub}
\end{eqnarray}
In contrast, near the opposite-phase fixed point, $\phi_0=\pi$, we get
\begin{eqnarray}
f_{x_\pi}=6a^3\phi'^3\int_{0}^{2\pi}\frac{g'(x)^3}{(1-2ag(x))^4}\left(\frac{1}{(1-2ag(x))}\frac{J_2}{J_3}\right)dx+O(\phi'^4),
\label{forcelubpi}
\end{eqnarray}
where we have defined
\begin{eqnarray}
J_n=\int_{0}^{2\pi}(1-2ag(x))^{-n}dx.
\end{eqnarray}
If we then assume $a\ll 1$ then Eq.~\eqref{forcelubpi} reduces to
\begin{eqnarray}
f_{x_\pi}\approx72a^4\phi'^3\int_{0}^{2\pi}g(x)g'(x)^3dx+O(\phi'^4).
\label{forcepilub}
\end{eqnarray}
We see then that for small amplitude waves, and small deviations in phase about the fixed points, the force about the in-phase configuration ($\phi_0=0$) is equal and opposite to the force about the  out-of-phase configuration ($\phi=\pi$). Unless both of them are neutrally stable (which is the case if the waveforms are too symmetric, see Sec.~\ref{sec:symmetry}) we therefore obtain the important result that, for a given waveform,  one fixed point will always be stable, while the other one will always be unstable. To determine which one is the stable point, one has to evaluate the geometric integral, $A=\int_{0}^{2\pi}gg'^3dx$. If $A<0$ then the fixed point at $\phi=0$ is stable, while it is the one at  $\phi=\pi$ in the case $A>0$. Stable passive hydrodynamic synchronization thus always  takes place for swimmers with asymmetric waveforms.

As a side note, we can also expand the force $\eqref{lubforce}$ in powers of $a\ll 1$, and we see that the leading order contribution is fourth order in amplitude, given for general $\phi$ as
\begin{eqnarray}
f_x\approx -36a^4\int_{0}^{2\pi}\Big(g(x+\phi)+g(x)\Big)\Big(g(x+\phi)-g(x)\Big)^3 dx,
\label{lubforcesmall}
\end{eqnarray}
plus terms at $O(a^6)$. We see that in the small amplitude limit there are only two fixed points for nontrivial waveforms $g$. The fourth-order scaling of the hydrodynamic force, Eqs.~\eqref{force0lub}, \eqref{forcepilub} and 
\eqref{lubforcesmall}, is reminiscent of the small-amplitude calculations from Sec.~\ref{smallamp} showing that no force can occur at second order in the wave amplitude, but a nonzero force does come at fourth order.

\subsection{Energy Dissipation}

The energy dissipated by viscous stress in the  volume $V$ of fluid between the sheets by is given by
\begin{eqnarray}
\dot{E}=\int_{V}\bsigma\boldsymbol{:}\nabla\bu \,dV.
\end{eqnarray}
In the lubrication limit, assuming unit width, the energy dissipation over one wavelength is then
\begin{eqnarray}
\dot{E}=\int_{0}^{2\pi}\int_{y_1}^{y_2}\left(\frac{\partial u}{\partial y}\right)^2dydx,
\end{eqnarray}
and given Eq.~\eqref{u} we have
\begin{eqnarray}
\dot{E}=\frac{1}{12}\int_{0}^{2\pi}h^3\left(\frac{dp}{dx}\right)^2dx,
\end{eqnarray}
which is explicitly
\begin{eqnarray}
\dot{E}=12\int_{0}^{2\pi}\frac{1}{h}\left(\frac{I_2}{I_3h}-1\right)^2dx.
\label{energylub}
\end{eqnarray}
We see the energy dissipation is non-negative and identically zero when $\phi=0$ (i.e. when $h$ is constant) and hence this must be a global minimum.

If we let $\phi=\phi_0+\phi'$ where $\phi'\ll 1$, we find near the in-phase conformation, $\phi_0=0$ 
\begin{eqnarray}
\dot{E}_0=12a^2\phi'^2\int_{0}^{2\pi}g'(x)^2 dx+O(\phi'^4),
\label{energy0}
\end{eqnarray}
and the energy increases quadratically with the slope of the wave from zero when $\phi=0$. Near the opposite-phase conformation, $\phi_0=\pi$, we get
\begin{eqnarray}
\dot{E}_\pi&=&12\left(J_1-\frac{J_2^2}{J_3}\right)-12\phi'^2\int_{0}^{2\pi}\Bigg\{\frac{g'(x)^2}{(1-2ag(x))^3}\nonumber \\
&&\times\left[1+\frac{6}{(1-2ag(x))}\frac{J_2}{J_3}\left(\frac{1}{(1-2ag(x))}\frac{J_2}{J_3}-1\right)\right]dx\Bigg\}+O(\phi'^4).
\end{eqnarray}
If we further assume that $a\ll 1$ we see that
\begin{eqnarray}
\dot{E}_\pi\approx 12a^2\int_{0}^{2\pi}\left[4g(x)^2-g'(x)^2\phi'^2\right]dx,
\label{energypi}
\end{eqnarray}
hence for any waveform $g(x)$ the energy dissipated between the sheets is maximum in the opposite-phase conformation, $\phi=\pi$.

Finally, if we expand the energy dissipation, Eq.~\eqref{energylub}, in small amplitude for general $\phi$, we find
\begin{eqnarray}
\dot{E}=12a^2\int_{0}^{2\pi}\big[g(x+\phi)-g(x)\big]^2 dx+O(a^3).
\label{energyluba}
\end{eqnarray}
We can see clearly again that the energy dissipation is a global minimum when $\phi=0$ and maximum when $\phi=\pi$ due to the $g(x+\pi)\rightarrow -g(x)$ symmetry of the waveform; this is in agreement with the previous small amplitude results for arbitrary separation.

An important consequence of the previous results is that, although the nature of the  fixed points depends on the swimmer waveform, the location of the minimum of energy dissipation does not. The conformation of minimum  energy dissipation is not necessarily stable: depending on the waveform geometry, the opposite-phase conformation, $\phi=\pi$, may be stable (specifically, when $A>0$) yet  it is the one corresponding to a  maximum of energy dissipation. 

Experimental evidence shows that  spermatozoa cells synchronize at the in-phase conformation (and indeed $A<0$ for the linearly increasing sine waves indicated by Rikmenspoel \cite{rikmenspoel65}). However, we find at least one instance, in the figures in Ref.~\cite{moore95}, which show spermatozoa flagella seemingly synchronized in opposite-phase (although no mention of phase difference is reported in the text).

\subsection{Dynamics}

After calculating the hydrodynamic force, we now look to solve for the time-evolution of the phase.  We thus assume that the sheets are force free, $f_x=0$, and find the corresponding value of $\UD$. From Eq.~\eqref{flowrate} we know
\begin{eqnarray}
\frac{\partial}{\partial x}\int_{y_1}^{y_2}u dy = \UD \frac{dy_2}{dx}\cdot
\end{eqnarray}
Integrating in $x$ and evaluating the integral in $y$ gives an expression for the pressure gradient as
\begin{eqnarray}
\frac{dp}{dx}=\frac{6\UD-12}{h^2}-\frac{12\UD y_2+C}{h^3},
\end{eqnarray}
where $C$ is a constant of integration. We find this constant by exploiting the periodicity of the pressure field, leading to
\begin{eqnarray}
C=\Big(6\UD(I_2-2K)-12I_2\Big)/I_3,
\end{eqnarray}
where $K=\int_{0}^{2\pi}y_2h^{-3} dx$.

The force on the upper sheet is given by
\begin{eqnarray}
f_x=\int_{0}^{2\pi}\left(\frac{1}{2}\frac{dp}{dx}(y_2+y_1)-\frac{\UD}{h}\right)dx.
\end{eqnarray}
We then solve for $\UD$ by enforcing that the sheets are force free. It is worth noting that when we set $\UD=0$, we retrieve the force from the static case given by Eq.~\eqref{lubforce}, which we label here $f_x^s$ to avoid confusion. Now since $\UD=-d\phi/dt$ we find that the phase evolves in time according to
\begin{eqnarray}
\frac{d\phi}{dt}=-\M f^s_x,
\label{dphilub}
\end{eqnarray}
where the mobility, $\M$, is given by
\begin{eqnarray}
\M^{-1} = \int_0^{2\pi} \left\{\frac{1}{h}-\left[ \frac{1}{h^2}- \frac{1}{h^3}\left(2y_2 + \frac{I_2-2K}{I_3} \right) \right] 3(y_2+y_1)\right\}\d x .
\end{eqnarray}
As physically expected, the rate at which the phase changes, Eq.~\eqref{dphilub}, is proportional to the static force, $f_x^s$, which would be applied if the sheets where not permitted to move. The result is a first-order integro-differential equation for $\phi$.

Expanding Eq.~\eqref{dphilub} for small amplitude, $a\ll 1$,  we find
\begin{eqnarray}
\frac{d\phi}{dt}\approx \frac{36a^4}{2\pi}\int_{0}^{2\pi}\Big(g(x+\phi)+g(x)\Big)\Big(g(x+\phi)-g(x)\Big)^3 dx,
\label{lubdynsmall}
\end{eqnarray}
plus terms at order $a^6$. We thus see that $d\phi/dt\sim -f_x^s/2\pi$ for small amplitude, and hence the mobility becomes  $1/2\pi$ in this limit. Notably, the result given by Eq.~\eqref{lubdynsmall} is the same as the one given by Eq.~\eqref{dphiexp} after proper dimensional rescaling.

We now expand near the fixed points by letting $\phi=\phi_0+\phi'$ and obtain
\begin{eqnarray}\label{newphi0}
\frac{d\phi'}{dt}\sim \pm \frac{36}{\pi}a^4A\phi'^3,
\end{eqnarray}
with a positive sign for $\phi_0=0$, and negative $\phi_0=\pi$. Solving this differential equation gives the exact phase dynamics for small times as
\begin{eqnarray}
\phi'=\frac{\text{sgn}(\phi'_i)}{\sqrt{{\phi}_i^{'-2}\mp 72a^4At/\pi}},
\label{phasespeedlub}
\end{eqnarray}
where $\phi'(t=0)=\phi'_i$. In the case of a stable fixed point, we thus get that the typical time  for synchronization scales as $t\sim 1/a^4|A|$, and thus the phase-locked state is reached faster for waves of larger amplitude ($a$ increases), and larger asymmetry ($|A|$ increases). Note that, as a difference, the typical time for synchronization in a viscoelastic fluids scales as the inverse square of the wave amplitude \cite{elfring10}.

\section{Boundary Integral Formulation}
\label{BIformulation}

The boundary integral method may be used to address numerically the synchronization between the swimmers  for shapes of arbitrary amplitude, as well as confirm our asymptotic results. We present in this section the principle of the method and our implementation of it, which is  quite similar to that given by Pozrikidis in his study of peristaltic pumping \cite{pozrikidis87b}, and hence  will be brief. The equations in the section are nondimensionalized similarly to what was done in Sec.~\ref{sec:definition}.

Consider any two solutions to the Stokes equations, $\{\bu, \bsigma\}$ and $\{\but,\bsigmat\}$ with no associated body forces  for any closed surface $S$ of outward normal $\bn$. The Lorentz reciprocal theorem  \cite{happel65} gives the equality
\begin{eqnarray}
\int_S \left(\bu\cdot\bsigmat-\but\cdot\bsigma\right)\cdot\bn \ dS=0.
\label{lorentz}
\end{eqnarray}
If we take for $\but$ and $\bsigmat$ in Eq.~\eqref{lorentz} the fundamental solutions for two-dimensional Stokes flow
\begin{eqnarray}
\but(\bx)&=&\frac{1}{4\pi}\bG(\hat{\bx})\cdot\tilde{\bf}(\bx_0), \\
\bsigmat(\bx)&=&\frac{1}{4\pi}\bT(\hat{\bx})\cdot\tilde{\bf}(\bx_0),
\end{eqnarray}
for the velocity and the stress at the field point $\bx$, due to the point force $\tilde{\mathbf{f}}$ at $\bx_0$, where $\hat{\bx}=\bx-\bx_0$ and the two-dimensional Stokeslet $\bG$ and stresslet $\bT$ are given by
\begin{eqnarray}
\bG&=&-\mathbf{I}\ln(\left|\bxh\right|)+\frac{\bxh\bxh}{\left|\bxh\right|^2},\\
\bT&=&-4\frac{\hat{\bx}\hat{\bx}\hat{\bx}}{\left|\hat{\bx}\right|^4},
\end{eqnarray}
then one obtains the boundary integral formulation of the two-dimensional Stokes equations for the velocity field within the fluid domain, $V$, and on the boundary, $S$, respectively,
\begin{eqnarray}
\bu(\bx_0)|_{\bx_0 \in V}&=&\frac{1}{4\pi}\int_S\left(\bu(\bx)\cdot\bT(\hat{\bx})\cdot\bn-\bb(\bx)\cdot\bG(\hat{\bx})\right)\ d S(\bx) \label{bi1},\\
\bu(\bx_0)|_{ \bx_0 \in S}&=&\frac{1}{2\pi}\int_S\left(\bu(\bx)\cdot\bT(\hat{\bx})\cdot\bn-\bb(\bx)\cdot\bG(\hat{\bx})\right)\ d S(\bx) \label{bi2},
\end{eqnarray}
where we have used $\bb=\bsigma\cdot\bn$.

Since the problem we consider is $2\pi$-periodic, we can reduce the domain of integration $S$ to a single period by using an infinite sum of periodically placed Stokeslets and stresslets,
\begin{eqnarray}
\bG^p&=&\sum_{n=-\infty}^{\infty}-\mathbf{I}\ln(\left|\hat{\bx}_n\right|)+\frac{\hat{\bx}_n\hat{\bx}_n}{\left|\hat{\bx}_n\right|^2},\\
\bT^p&=&\sum_{n=-\infty}^{\infty}-4\frac{\hat{\bx}_n\hat{\bx}_n\hat{\bx}_n}{\left|\hat{\bx}_n\right|^4},
\end{eqnarray}
where $\hat{\bx}_n=\{\hat{x}_0+2\pi n,\hat{y}_0\}$ \cite{pozrikidis92}. As shown in Ref.~\cite{pozrikidis87} these may be expressed in closed form by using the summation formula
\begin{eqnarray}
B=\sum_{n=-\infty}^{\infty}\ln(\left|\hat{\bx}_n\right|)=\frac{1}{2}\ln\left[2\cosh(\hat{y}_0)-2\cos(\hat{x}_0)\right].
\end{eqnarray}
We can then construct the elements of $\bG^p$ and $\bT^p$ using $B$ and its derivatives as follows:
\begin{eqnarray}
G_{xx}^p &=& -B-\partial_yB+1,\nonumber\\
G_{xy}^p &=& y\partial_xB,\nonumber\\
G_{yy}^p &=& -B+y\partial_yB,\nonumber\\
T^p_{xxx} &=& -2(2\partial_xB+y\partial_{xy}B),\nonumber\\
T^p_{xxy} &=& -2(\partial_yB+y\partial_{yy}B),\nonumber\\
T^p_{xyy} &=& 2y\partial_{xy}B,\nonumber\\
T^p_{yyy} &=& -2(\partial_yB-y\partial_{yy}B),
\end{eqnarray}
where the Stokeslet and stresslet are invariant under permutation of its indices \cite{pozrikidis92}.

Following the approach outlined by Higdon \cite{higdon85}, the boundary $S$ (the surface of each sheet as the sides cancel) is discretized into $2N$ straight line elements $S_n$. We assume the stress $\bb$ and the velocity $\bu$ are linear functions over a particular interval  ($\bb\rightarrow\bb_n$, $\bu\rightarrow\bu_n$) and then collocate $\bx_0$ at each of the $2N$ segments, $\bx_0\rightarrow \bx_m$, to obtain a system of $N$ equations with $N$ unknowns. The periodic Stokeslet and stresslet are regularized by subtracting off from them their non-periodic counterparts and then adding back the difference; the two-dimensional Stokeslet and stresslet are then integrated analytically. We hence have
\begin{eqnarray}
\bu(\bx_m)&=\displaystyle \frac{1}{2\pi}\sum_{n=1}^{2N}\Bigg[&\int_{S_n}\bu_n\cdot\left(\bT^p-\bT\right)\cdot\bn_n d S_n-\int_{S_n}\bb_n\cdot\left(\bG^p-\bG\right)d S_n \nonumber\\
&&+\int_{S_n}\bu_n\cdot\bT\cdot\bn_n dS_n-\int_{S_n}\bb_n\cdot\bG dS_n\Bigg].
\end{eqnarray}
The regularized integrals have a removable singularity at $\bx=\bx_m$ which is obtained by Taylor expansion. The boundary integral formulation is thereby reduced to a linear system that can be inverted using standard techniques to obtain the stress $\bb$. The force on the top sheet is then given by integrating the stress
\begin{eqnarray}
f_x=\sum_{n=N+1}^{2N} \left[\be_x\cdot \int_{S_n} \bb_n dS_n\right].
\end{eqnarray}

In order to solve for the dynamical problem we let the boundary condition be represented by a sum of a surface deformations and an unknown rigid body motion, $\bu\rightarrow \bu_n+\UD\be_x$, on the upper sheet. The additional unknown, $ \UD$,  is found by enforcing that the sheets are force free, $f_x=0$.

The numerical procedure was validated through convergence tests and by reproducing previous results for shear flow over sinusoidal surfaces \cite{pozrikidis87}.

\section{Results}\label{results}

\begin{figure}
\includegraphics[width=0.35\textwidth]{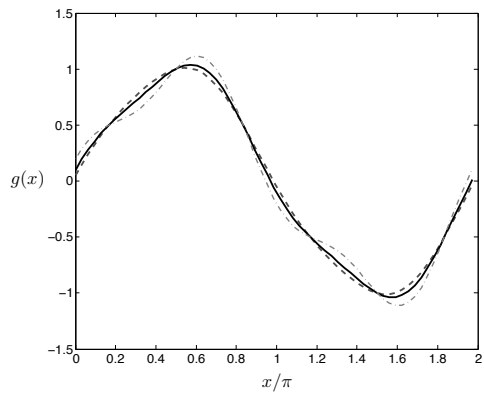}
\caption{Illustration of the waveform $g(x)=\sin x+\gamma\cos 3x$ for varying asymmetry. The dashed line corresponds to $\gamma=0.05$, solid line $\gamma=0.1$, and dash-dot line $\gamma=0.2$.}
\label{shapes}
\end{figure}

We now present in this section the  results of both our asymptotic and numerical calculations to address the synchronization of  specific waveforms. For illustrative purposes we restrict ourselves here to the family of waveforms described by the function,
\begin{eqnarray}
g(x)=\sin x+\gamma\cos 3 x,
\label{waveform}
\end{eqnarray}
i.e. $\beta_1=1$, $\alpha_3=\gamma$, and all other modes equal to zero, as illustrated in Fig.~\ref{shapes}. In essence these shapes are  geometric perturbations (small for small $\gamma$) to Taylor's sinusoidal swimming sheet. They have a broken front-back symmetry when $\gamma$ is nonzero. Reversing the sign of $\gamma$ is equivalent to reflecting the geometry of each wave about the vertical axis, $\{\gamma\rightarrow -\gamma\} = \{x \rightarrow -x+\pi\}$, which itself is equivalent to reversing the kinematics of the problem. In other words, changing the sign of $\gamma$ changes the sign of the forces on the sheets which leads to stable fixed points becoming unstable, and vice versa. In addition, the simple form of $g(x)$ allows us to obtain some explicit formulae from the general theory presented in Secs.~\ref{smallamp} and \ref{lublimit}.

In the lubrication limit, the geometric parameter $A=\int_{0}^{2\pi}gg'^3dx=-2\pi\gamma$ controls the evolution of the phase near fixed points. We see that $\gamma > 0$ gives $A<0$,  which leads a stable fixed point at $\phi=0$ and unstable at $\phi=\pi$. By  symmetry,  $\gamma < 0$ necessarily gives $A>0$,  and thereby exchanges the location of the stable and unstable fixed points. In addition,  from Eq.~\eqref{lubforcesmall}, we have that when $a\ll 1$ the phase locking force is given by
\begin{eqnarray}
f_x=144\pi a^4 \gamma\sin^3\phi,
\end{eqnarray}
which  is linear in the asymmetry and quartic in the amplitude, and leads to a time-evolution of the phase as
\begin{eqnarray}
\frac{d\phi}{dt}=-72a^4\gamma\sin^3\phi.
\end{eqnarray}

The energy dissipation in the lubrication limit, for $a\ll 1$, Eq.~\eqref{energyluba}, is
\begin{eqnarray}
\dot{E}\approx 24\pi a^2\left[1-\cos\phi+\gamma^2(1-\cos 3\phi)\right].
\label{energylub3}
\end{eqnarray}
Similarly, in the small amplitude limit, Eq.~\eqref{energyexp} yields
\begin{eqnarray}
\dot{E}\approx 2\pi \epsilon^2 \Big[A(\hb)-B(\hb)\cos\phi+3^3\gamma^2\big(A(3\hb)-B(3\hb)\cos3\phi\big)\Big],
\label{energyexp3}
\end{eqnarray}
where the functions $A$ and $B$, given by Eq.~\eqref{AB}, introduce a dependance on the separation $h$, and  Eq.~\eqref{energyexp3} reduces to Eq.~\eqref{energylub3} when $\hb$ is small (after accounting for the separate scalings).

We see clearly that the energy dissipation rate is invariant under $\gamma\rightarrow-\gamma$ and further, when  we are assuming that $\gamma$ is a small, the change in the energy dissipation due to the asymmetry is also small, $O(\gamma^2)$.

\subsection{Comparison between asymptotic and numerical methods}

\begin{figure}[t]
\includegraphics[width=0.35\textwidth]{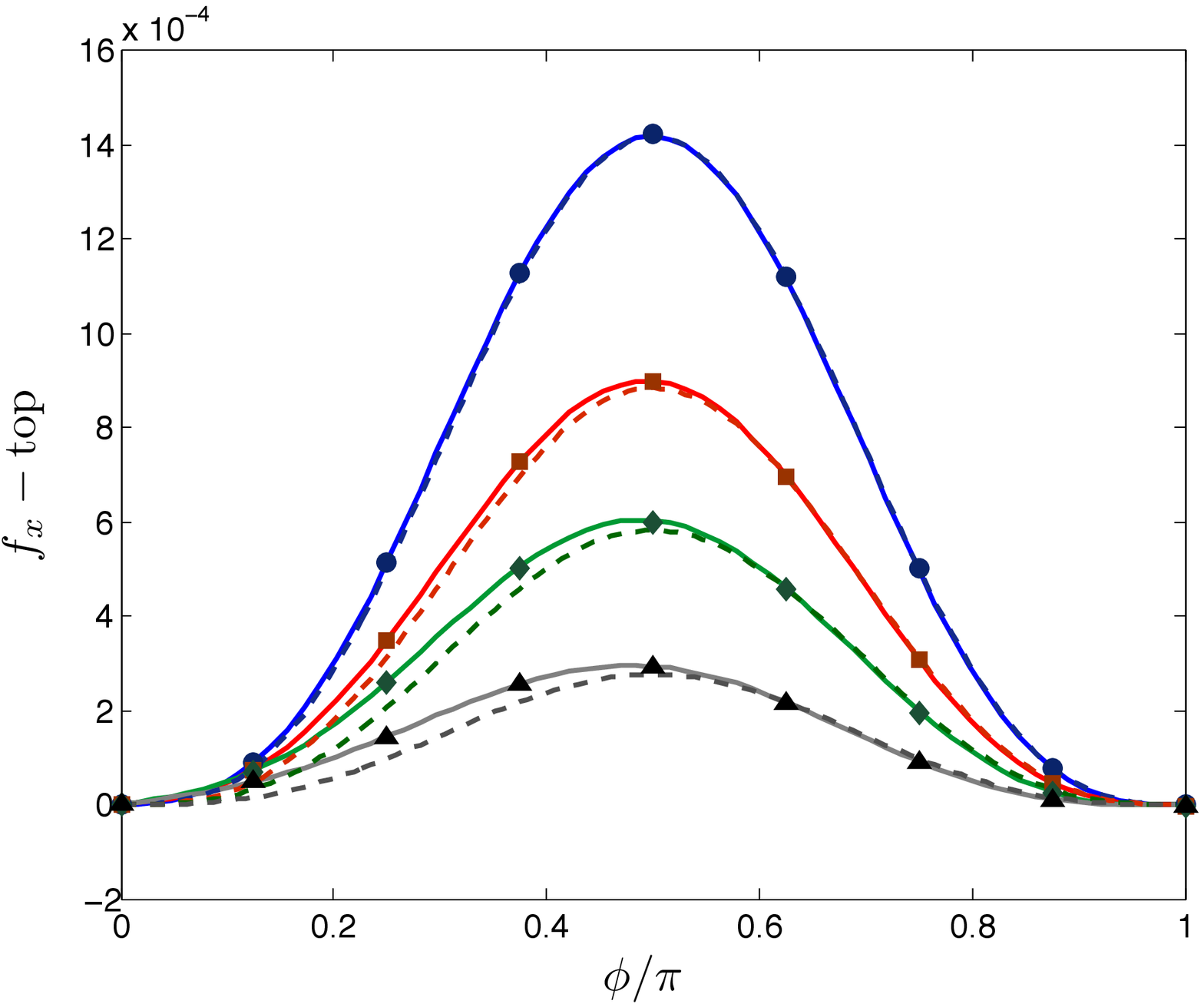}\\
\includegraphics[width=0.35\textwidth]{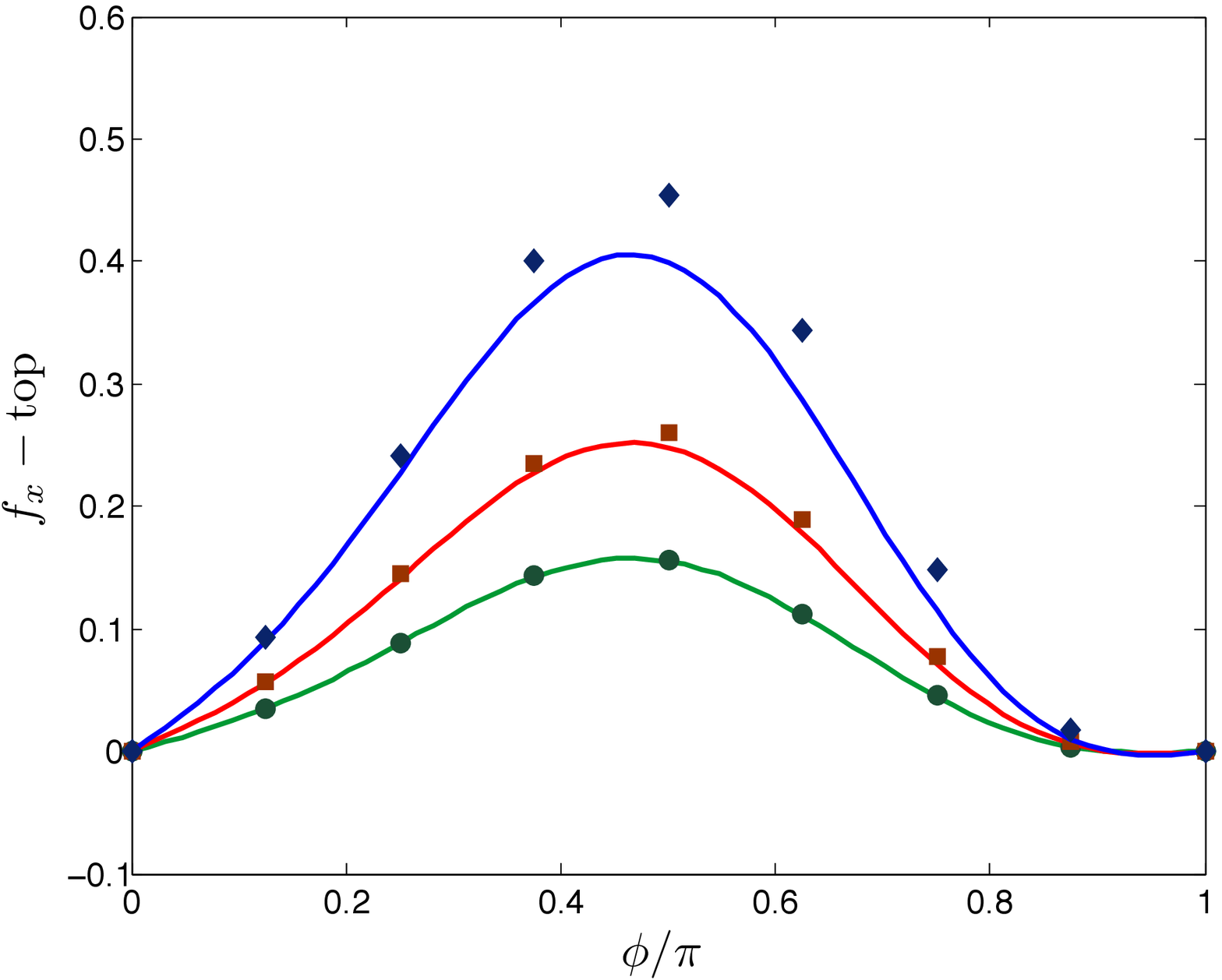}
\caption{(Color online). Force, $f_x$, vs. phase difference,  $\phi$, for an asymmetry of $\gamma=0.1$, in  the lubrication limit (dashed lines, top figure only), small amplitude limit (solid lines, both figures), and using the boundary integral method (symbols, both figures).
Top: fixed amplitude, $\epsilon=0.01$, and  varying swimmer-swimmer distances, $\bar h$; blue circles: $f_x$, $\hb=0.2$; red squares: $10f_x$, $\hb=0.4$; green diamonds: $f_x/\epsilon$, $\hb=0.6$; gray triangles: $2f_x/\epsilon$, $\hb=0.8$. 
Bottom: Fixed separation distance $\hb=1$ and varying waveform amplitudes; green circles: $\pi f_x/\epsilon$, $\epsilon=0.1$; red squares: $\pi f_x$ with $\epsilon=0.2$; blue diamonds: $f_x$, $\epsilon=0.4$. 
We observe a gradual breakdown of the lubrication approximation for increased separation, $\bar h$ (top), and  of the small amplitude expansion for increased amplitude $\epsilon$ (bottom). Note that forces have been scaled for display purposes.}
\label{lubexpbreak}
\end{figure}

In the small amplitude limit (\ref{smallamp}) we have explicitly assumed that $\epsilon \ll 1$, and also implicitly that $\epsilon \ll \hb$. The lubrication limit (\ref{lublimit}) effectively captures the physics of the problem when the sheets are quite close together, i.e. the limit $\hb \ll 2\pi$. If  we  want in addition  the phase, $\phi$, to be able to span the range of all possible values then we also get the geometrical constraint $\epsilon < \hb/2$ (or, in terms of lubrication variables, $a < 1/2$). There exists therefore a regime in which both asymptotic approaches  are valid, namely the limit $\epsilon \ll h\ll 1$.

As a validation of our methods we plot the analytical results from both asymptotic limits, together with the numerical results, for such a regime in Fig.~\ref{lubexpbreak} (top). The static force on the top sheet, $f_x^s$, is shown for the waveform of Eq.~\eqref{waveform} with an asymmetry of $\gamma=0.1$ and wave amplitude $\epsilon=0.01$  (in this plot the forces have been scaled for display purposes only, see caption).  The solid lines represent the small amplitude limit, dashed lines the lubrication limit, and symbols are for the numerical data obtained from the boundary integral method. The results from all three methods agree quantitatively for small swimmer-swimmer separation, $\hb$. As the value of $\hb$ increases, the lubrication results start to deviate, but the small amplitude results remains accurate (recall that $\epsilon \ll 1$ in all cases).

For larger values of the separation distance between the swimmers, the lubrication results cannot be applied, but the small-amplitude asymptotics, Eq.~\eqref{f4}, remain valid as long as the wave amplitude is small. The value of the static force is compared to the numerical results  in Fig.~\ref{lubexpbreak} (bottom) for large separation, $\hb=1$, and as function of the wave amplitude, $\epsilon$. The agreement between the two is excellent for $\epsilon=0.1$, but  they deviate quantitatively for larger wave amplitudes (although the correct order of magnitude, and dependence on $\phi$, is obtained).

\subsection{Stability}

\begin{figure}[t]
\includegraphics[width=0.35\textwidth]{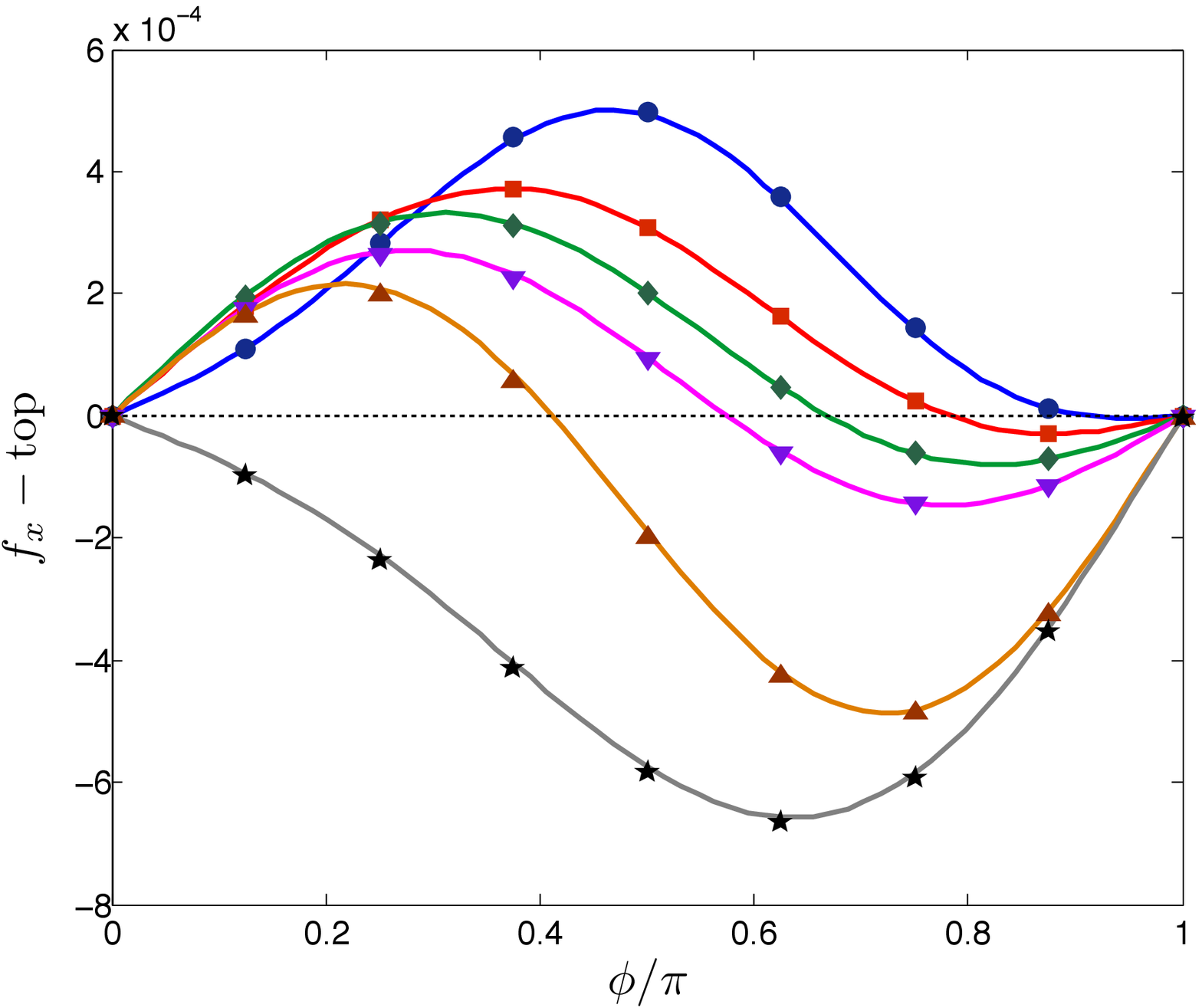}\\
\includegraphics[width=0.36\textwidth]{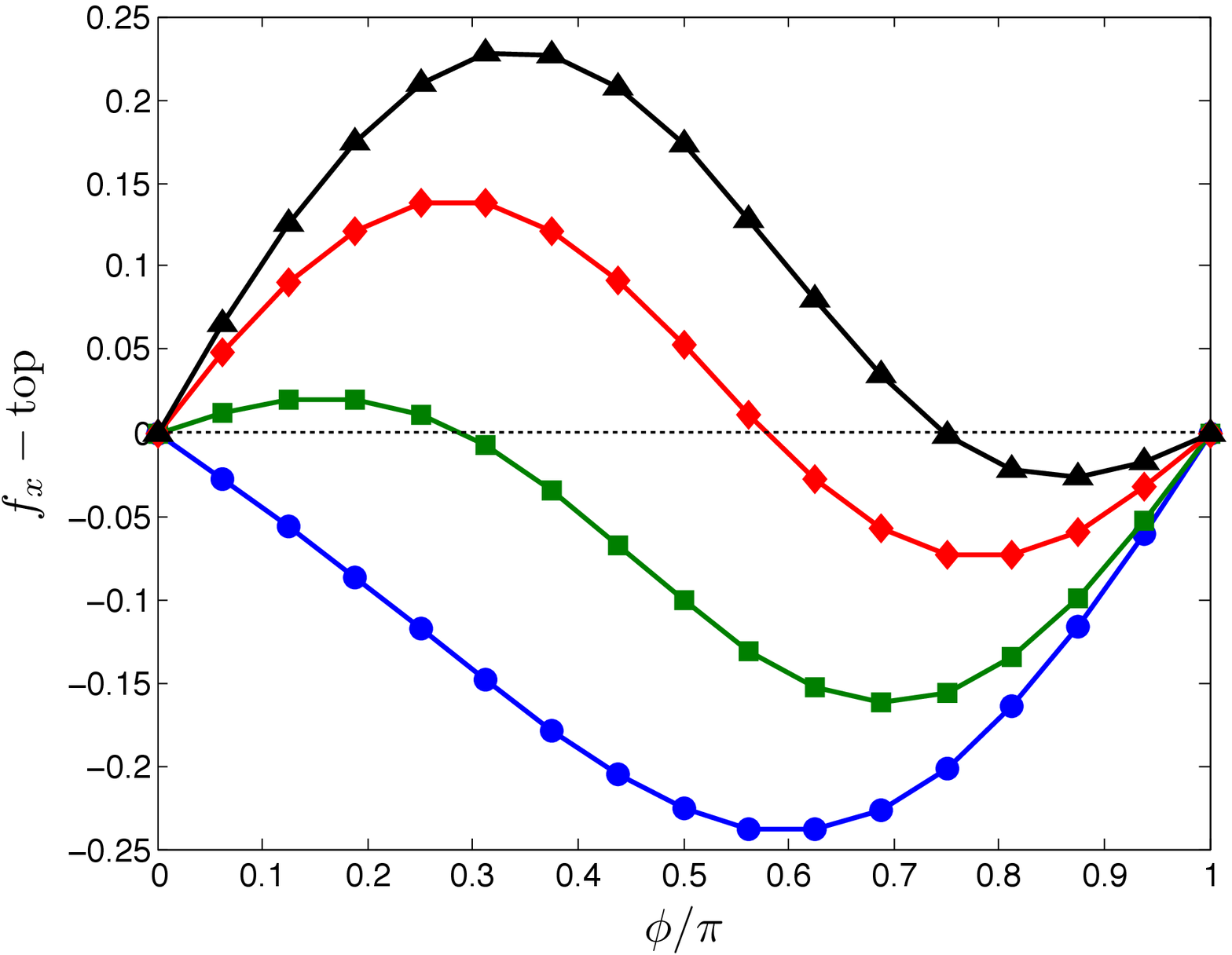}
\caption{(Color online). Dependence  of the force, $f_x$, on the phase $\phi$ for varying separation $\hb$ with an asymmetry $\gamma=0.1$. 
Top: small dimensionless amplitude, $\epsilon=0.1$. The solid lines are obtained in the small amplitude limit while symbols are for boundary integral computations; blue circles: $\epsilon f_x$, $\hb=1$; red squares: $2f_x$, $\hb=2$; green diamonds: $\pi f_x/2\epsilon$, $\hb=3$; 
purple down triangles: $f_x/\epsilon^2$, $\hb=4$; 
orange up triangles: $f_x/\epsilon^3$, $\hb=5$; gray stars: $f_x/\pi\epsilon^4$, $\hb=6$. 
Increasing the distance between the sheets introduces an additional fixed point  not present in the lubrication limit, and its position  moves with $\hb$.
Bottom: numerical results using the boundary integral method (solid line and symbols) in the case of high amplitude waves, $\epsilon=1$. Black triangles: $f_x$, $\hb=3$; red diamonds: $2\pi f_x$, $\hb=4$; green squares: $10\pi f_x$, $\hb=5$; blue circles: $100f_x$, $\hb=6$. Forces have been scaled for display purposes.
}
\label{fixedpoint_height}
\end{figure}

\begin{figure}
\includegraphics[width=0.35\textwidth]{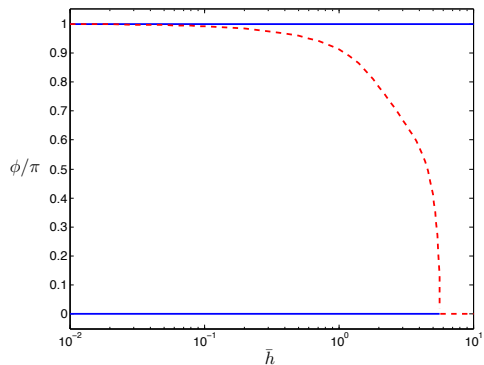}
\caption{(Color online). Location of the  fixed points  (value of the phase $\phi$ such that $f_x(\phi)=0$) as a function  of $\hb$ as obtained in the small amplitude limit for $\gamma>0$. The solid blue lines indicate a stable fixed point whereas the dashed red line indicates an unstable fixed point. The position of the unstable fixed point moves with $\hb$: it is created for small values of $\hb$ near $\phi=\pi$,  migrates to the left, and merges with $\phi=0$ at the critical value $\hb\approx 5.65$. In the opposite case where $\gamma<0$, the stable fixed points become unstable and vice versa.}
\label{stability}
\end{figure}

When we introduce a variable separation between the swimmers, $\hb$, and thus go beyond the small $\hb$ limit from the lubrication approximation, we  get  that the number of fixed points and their  nature does not depend solely on the waveform geometry, but actually also on the swimmer-swimmer distance. 
In Fig.~\ref{fixedpoint_height} (top) we show the dependence of the static force on the phase,  
for an amplitude  $\epsilon=0.1$  and an asymmetry $\gamma=0.1$, as we vary the separation between the swimmers $\hb$ (line:  small-amplitude asymptotics; symbols:  boundary integral computations). 
A fixed point is a conformation with  phase difference $\phi$ such that $f_x(\phi)=0$;  if the slope of the force is positive the fixed point is stable, while a negative slope indicates an unstable fixed point. What we see in Fig.~\ref{fixedpoint_height} (top) is that increasing the separation between the sheets from the small $\hb$ values in the lubrication limit gives rise to an additional fixed point. In the case illustrated in Fig.~\ref{fixedpoint_height} (top), this new fixed point is always unstable. It first appears  near $\phi=\pi$ (leading to the fixed point at $\phi=\pi$ becoming stable),  moves toward $\phi=0$ when the separation distance between the swimmers increases, and eventually merges with $\phi=0$, which then turns to an unstable point, at a critical value of $\hb$.

\begin{figure}
\includegraphics[width=0.35\textwidth]{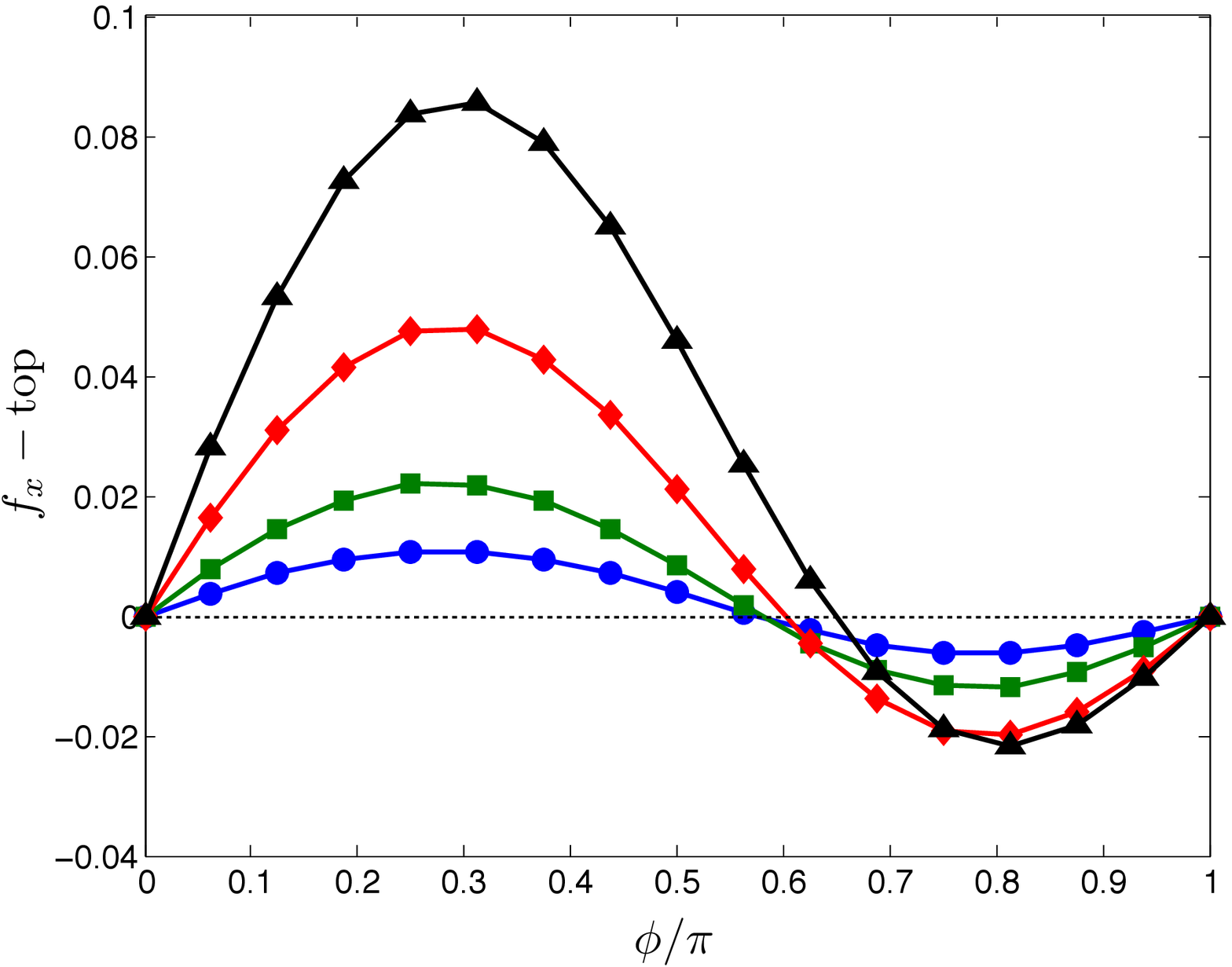}\\
\includegraphics[width=0.35\textwidth]{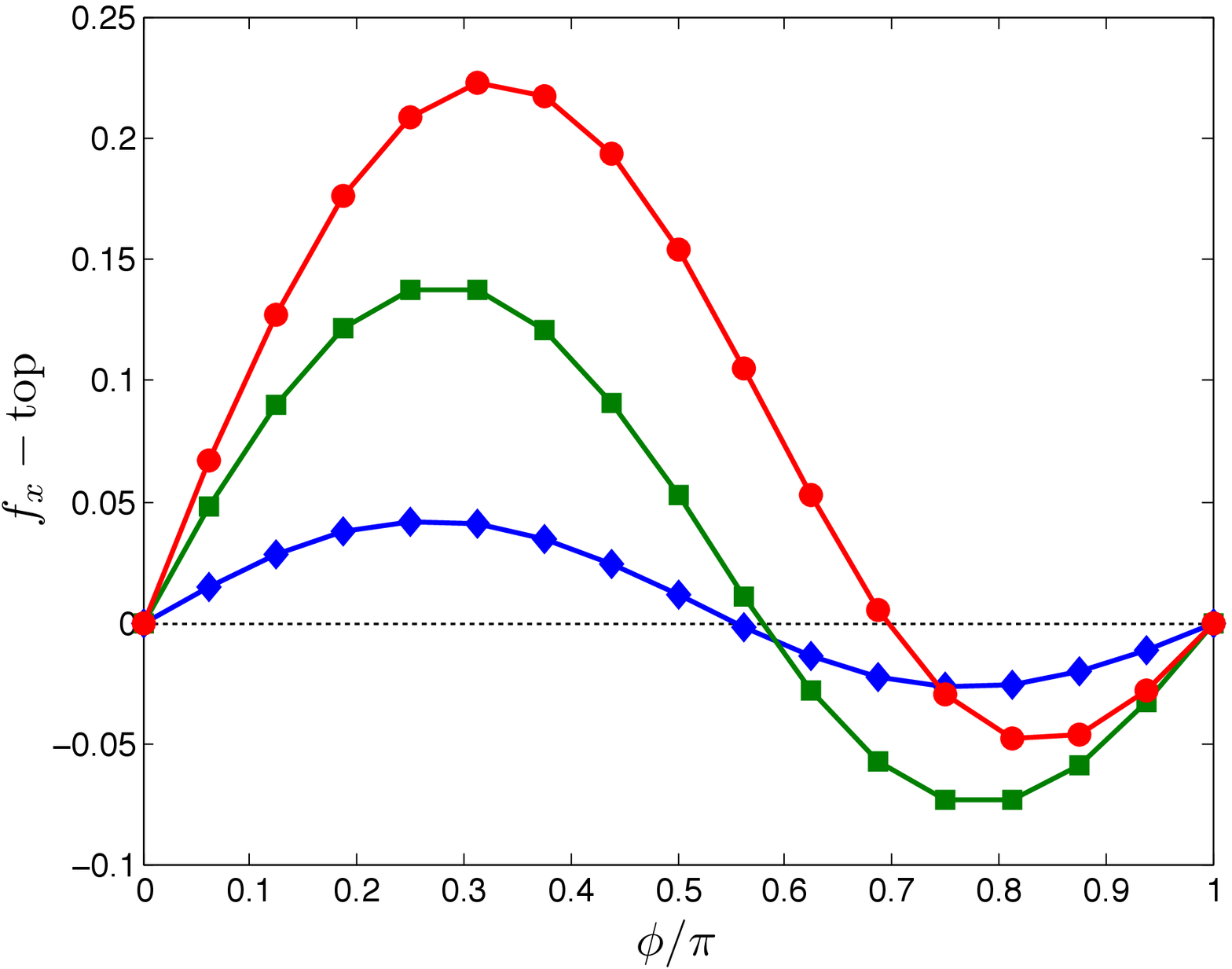}
\caption{(Color online). Plot of the force on the top swimmer, $f_x$, as a function of the phase difference, $\phi$, using the boundary integral method with $\hb=4$. Top: $\epsilon=1$ for varying asymmetry; blue circles: $\gamma=0.05$; green squares: $\gamma=0.1$; red diamonds: $\gamma=0.2$; black triangles $\gamma=0.3$. We see that for large amplitude waves the force is no longer linear with asymmetry as evidenced by the moving of the middle fixed point.
Bottom: $\gamma=0.1$ for varying large amplitude waves; blue diamonds: $10\pi f_x$, $\epsilon=0.5$; green squares: $2\pi f_x$, $\epsilon=1$; red circles: $f_x$, $\epsilon=1.5$. We see that for $\epsilon \le 1$ the location of the middle fixed point remains close to the small amplitude limit, while it has drifted significantly for $\epsilon=1.5$. Forces have been scaled for display purposes.}
\label{gamma_amplitude}
\end{figure}

In Fig.~\ref{stability} we display the location of the fixed points explicitly as a function of $\hb$ in the small amplitude limit for $\gamma>0$. In this limit, the force is linear in the asymmetry, $\gamma$, therefore the location of the fixed points is invariant under a linear scaling of the asymmetry, $\gamma\rightarrow b\gamma$ where $b>0$, while the nature of the fixed points changes with a change of the sign of $\gamma$. The appearance of a new fixed point, described in the previous paragraph, is apparent. As $\hb$ tends asymptotically to zero, $\phi=0$ is stable (blue solid line) while $\phi=\pi$ is unstable (red dashed line), which is the lubrication result. For intermediate values of $\hb$, both 0 and $\pi$ are stable, and the new fixed point moves from $\pi$ to 0 as $\hb$ increases. It merges with $\phi=0$ for a critical distance between the swimmers ($\hb\approx 5.65$ for our choice of waveform), at which point $\phi=0$ becomes  unstable, and remains so for larger values of $\hb$. As expected, upon a reversing the sign of $\gamma$, stable fixed points become unstable and vice versa.

Further analysis of the equations of motion shows that the additional fixed point that arises when $\hb$ is past the lubrication limit is a direct consequence of the inextensible boundary conditions. In the lubrication limit, the boundary conditions are extensible insofar as the there is only a vertical component, however away from this limit there arises horizontal motion to maintain inextensibility, and it is precisely this horizontal motion which leads to the additional dynamic complexity. Conversely for extensible boundary conditions, Eq.~\eqref{ext}, the fixed points remain unchanged from those in the lubrication limit.

Using the boundary integral formulation it is possible to extend these results to large amplitude waves. In Fig.~\ref{fixedpoint_height} (bottom) we show the horizontal force on the upper sheet as a function of the phase between the swimmers for various mean swimmer-swimmer separation but now with $\epsilon=1$. The results are qualitatively similar to those obtained in the  small amplitude limit, with the occurrence of a new fixed point, unstable, and moving from $\phi=\pi$ to $\phi=0$ as the separation increases.
A difference  we do observe between small and large amplitude is that the location of the fixed point is no longer invariant under a change in the  asymmetry factor, $\gamma$. In Fig.~\ref{gamma_amplitude} (top) we show that an increase in the asymmetry factor leads to a small, but nonzero, migration of the mobile fixed point toward $\pi$. A similar drift  is obtained with an  increase in the waveform amplitude  (Fig.~\ref{gamma_amplitude}, bottom).


\subsection{Dynamics}

\begin{figure}[b]
\includegraphics[width=0.35\textwidth]{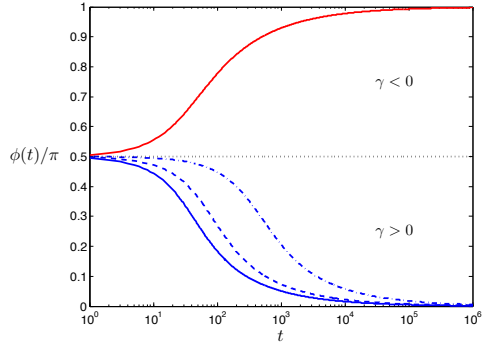}
\caption{(Color online). Time-evolution of the phase, from $\phi_0=\pi/2$, in the lubrication limit with $\delta = 0.1$. The dashed line indicates $\gamma=0$ while the solid line indicates $\gamma=\pm 0.1$, $a=1/4$. The dashed line has $\gamma=0.05$, $a=1/4$ and the dash-dot line has $\gamma=0.1$, $a=1/8$.}
\label{dynamicslub}
\end{figure}

The time-evolution of the phase   is given in general by the integro-differential equation
\begin{eqnarray}\label{dyna}
\frac{d\phi}{dt}=-\M(\phi) f_x^s(\phi).
\end{eqnarray}
As noted above,  in the small amplitude limit the mobility becomes independent of the phase, $\M=\hb/2\pi$, hence in that case the dynamics is completely set by the static force. Note that the mobility is never zero so no additional fixed points arise from Eq.~\eqref{dyna}.

In the lubrication limit we know that there exist only two fixed points, and the location of the stable fixed point depends only on the  waveform asymmetry. In Fig.~\ref{dynamicslub} we plot the time-evolution of the phase in this limit. We see that if the system is symmetric ($\gamma=0$), indicated by the black solid line, then the phase remains constant in time. This corresponds to the no-synchronization situation discussed in Sec.~\ref{sec:symmetry}. When we introduce an asymmetry, $\gamma \ne 0$, then the two swimmers phase lock over time. When $\gamma > 0$ then $A < 0$ and the system evolves to a stable in-phase conformation, and opposite-phase for the converse. Given that the amplitude, $a$, is reasonably small for all curves (we have the geometrical constraint $a<1/2$), we observe roughly the same dependance of the typical time scale  for phase locking, $t$, on the wave asymmetry and amplitude as in the small-amplitude limit  (for which $t\sim a^{-4}\gamma^{-1}$, see Eq.~\ref{newphi0}).

\begin{figure}
\includegraphics[width=0.35\textwidth]{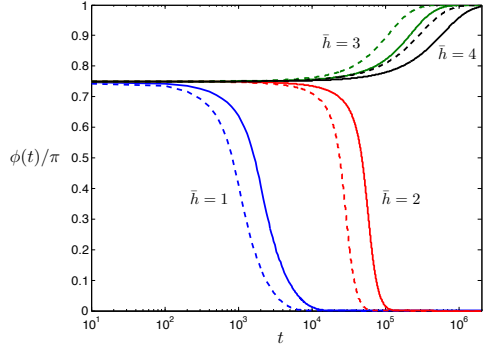}
\caption{(Color online). Time-evolution of the phase, from the initial condition $\phi_0=3\pi/4$, for $\hb=[1\to4]$ in the small amplitude limit with $\epsilon = 0.1$. A solid line indicates $\gamma=0.1$ while a dashed line indicates $\gamma=0.2$.}
\label{dynamicsexp}
\end{figure}

We have seen above  that with an increase in $\hb$ comes additional  fixed point, and thus we expect the phase dynamics to depend similarly on $\hb$. In Fig.~\ref{dynamicsexp} we plot the time-evolution of the phase in the small amplitude limit for various values of the swimmer-swimmer distance  in the case where $\gamma>0$. Given that the phase mobility, $\M$, is independent of the asymmetry, and that the force is linear in $\gamma$, we find that the time scale for synchronization  scales with the inverse of the asymmetry factor, i.e. $t\sim \gamma^{-1}$, as it does when $\hb\ll 1$. The final stable swimmer-swimmer conformation can be understood simply by recalling the force plot in Fig.~\ref{stability}. If the initial conformation is to the left of the moving unstable point, then the sheets evolve to $\phi=0$, while they start to the right they evolve to $\phi=\pi$. If we reverse the asymmetry of the waveforms, $\gamma < 0$, then the converse is true, the fixed point which varies with separation represents the only stable conformation for intermediate values of $\hb$ and we obtain synchronization to a fixed finite phase difference, $0<\phi<\pi$, as is observed in the metachronal beating of cilia.

\begin{figure}[b]
\includegraphics[width=0.35\textwidth]{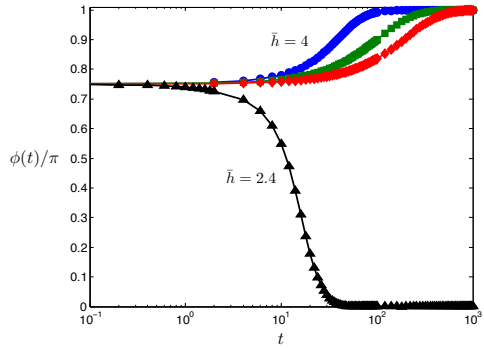}
\caption{(Color online). Time-evolution of the phase, from $\phi_0=3\pi/4$, for large amplitude waves using the boundary integral method. The blue circles indicate $\hb=4$, $\gamma=0.1$, $\epsilon=1.5$; green squares: $\hb=4$, $\gamma=0.1$, $\epsilon=1$; red diamonds: $\hb=4$, $\gamma=0.05$, $\epsilon=1$ and black triangles: $\hb=2.4$, $\gamma=0.1$ and $\epsilon=1$.}
\label{dynamicsLarge}
\end{figure}

A similar plot is shown in Fig.~\ref{dynamicsLarge}  in the case of large amplitude waves, using the boundary integral method,  starting from an initial relative phase of $\phi=3\pi/4$ and with a positive asymmetry, $\gamma >0$. Again the essential physics is well captured by the small amplitude expansion: there exists a critical swimmer-swimmer separation below which the sheets evolve to the in-phase conformation. This is seen  in Fig.~\ref{dynamicsLarge} where with $\hb=2.4$, $\epsilon=1$ and $\gamma=0.1$  the sheets evolve to $\phi=0$ (in phase) whereas when the distance is increased to $\hb=4$ the sheets evolve to $\phi=\pi$ (opposite phase). A waveform with a larger amplitude, $\epsilon=1.5$, leads to a faster evolution of the phase than for $\epsilon=1$  for equal asymmetry ($\gamma=0.1$), which in turn evolves faster than for equal amplitude, $\epsilon=1$, but lower asymmetry $\gamma=0.05$. We note however that for large amplitude waves, the effect of increasing the amplitude on the rate of phase change is drastically reduced; in the small-amplitude limit the rate of evolution is quartic with the wave amplitude and here we see an effect which is less than cubic. Despite the reduction, the effect of amplitude is still strong and we observe drastically faster synchronization for order one amplitudes.

\begin{figure}[t]
\includegraphics[width=0.8\textwidth]{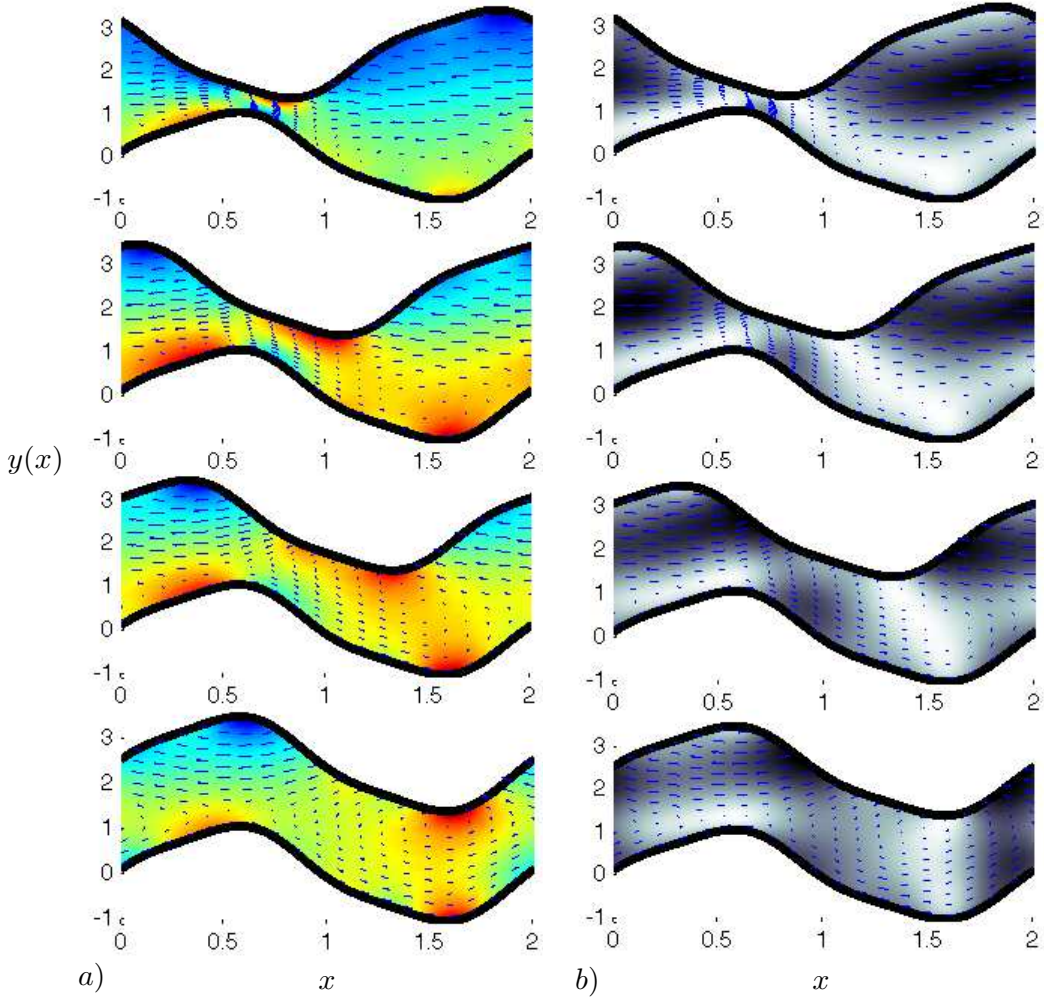}
\caption{(Color online). Illustration of the flow field during synchronization for $\epsilon=1$, $\gamma=0.1$ and $\hb=4$ using boundary integral computations; (a) flow vorticity;  (b) iso values of $|\bu|^2$. Darker regions correspond to higher velocity and vorticity, and arrows indicate instantaneous velocity vector field. The two sheets start from the initial condition $\phi_0=3\pi/4$ (top panel). Time increases top to bottom with the phase $\phi$ being equal to $\{3\pi/4,\pi/2,\pi/4,0\}$ in the four panels. 
}
\label{fieldb}
\end{figure}

To illustrate the flow  driving the synchronizing dynamics shown in Fig.~\ref{dynamicsLarge}, we produce snapshots of the flow field for the $\hb=2.4$, $\epsilon=1$, $\gamma=0.1$ conformation in Fig.~\ref{fieldb}. We display the out-of-plane vorticity, $\omega$, in Fig.~\ref{fieldb}a  and the squared velocity field, $|\bu|^2$, in Fig.~\ref{fieldb}b. Both plots are overlaid with arrows indicating the velocity vector field. The darker regions indicate higher vorticity and velocity in each plot respectively. Time increases from top to bottom, and we show the instances where the phase angle is given by $\phi=\{3\pi/4,\pi/2,\pi/4,0\}$, corresponding to relative velocity of the sheets, $\UD=\{0.0325,0.1223,0.1097,0\}$.

\section{Discussion}\label{conclusion}

\subsection{Summary of results}

In this paper we have considered, as a model for the synchronization of flagellated cells, the relative motion of two free-swimming planar parallel sheets propagating waves of transverse displacement. We showed that due to the kinematic reversibility of the Stokes equations, waveform conformations with both vertical and horizontal axis symmetry could not yield synchronizing dynamics. When we break vertical axis symmetry, the phase of the system evolves to stable dynamic equilibria whose position is entirely dependent on the geometry of flagellar waveforms, and the distance between them. 

When the swimmers are close together we are able to make use of the lubrication equations and find two fixed points, in-phase and opposite-phase. The location of the stable point depends on the nature of the asymmetry, which is measured by an integral over the waveform geometry. If the front-back asymmetry of the geometry is reversed it is equivalent to reversing the kinematics of the problem which yields a reversal of forces. In other words, stable points become unstable and vice versa. In contrast, the energy dissipation is always a minimum for the in-phase conformation, and indicates the possibility of phase-locking into a conformation of maximum energy dissipation.

An expansion in small amplitude is utilized to introduce inextensible boundary conditions and order one distances between the organisms. In this case there arise additional fixed points, whose location and nature depends on the cells geometry and separation. Among the possibilities is synchronization at a stable intermediate phase between in-phase and opposite-phase.

Finally, we presented numerical results for large-amplitude waves using the boundary integral method. The computational results indicate that between the lubrication limit and the small amplitude expansion, all the relevant physics can be captured analytically. However, since  the phase locking force depends so strongly on waveform amplitude, we observe much faster synchronization for large amplitudes, as might be expected.

\subsection{Two-dimensional modeling and collective locomotion}

The two dimensional model used here is admittedly too simple to provide quantitative predictions for real microorganisms. However the simplicity allows analytic formulae to be derived and a mathematical structure elucidating the interaction between the bodies due to the Stokesian flow to be obtained, and  gives an explicit description of the effect of symmetry breaking. The intuition garnered here may then be useful for more complex models, with finite three-dimensional bodies, that must be solved entirely numerically.

We first note that, as a result  of the  two-dimensional approach, the viscous mobility of the cells in the direction perpendicular to that of the wave propagation is strictly equal to zero. For real microorganisms however this is not the case and hence fluid forces will determine the separation between the swimmers dynamically. Since swimming cells are force-free, the far-field velocity is typically a force dipole. In particular spermatozoa have a positive force dipole (so-called pushers \cite{lauga09b}). Far-field interactions between pushers tend to both align the major axis of pushers drive them together. Accordingly experimental evidence suggests that as spermatozoa synchronize they come together very tightly \cite{woolley09} (see also Fig.~\ref{sync}). In light of this, the lubrication limit is perhaps the relevant physical limit to consider for the phase locking of swimming cells. In contrast, eukaryotic cilia are attached to a substrate at a fixed distance which is varies depending on the organism \cite{brennen77}.

In addition, the two-dimensional limit offers one particularity, which is that the fluid between the swimmers  (inner problem) does not communicate with that outside the swimmers (outer problem).  The outer problem, that addressed by Taylor, is purely kinematic, in the sense that the speed of the sheet relative to the flow at infinity is uniquely determined without resorting to a balance of forces, unlike the inner problem. Further to this, because the outer problem cannot impose a force on the outer surface of the sheet, the forces are individually zero for the inner problem and therefore the inner problem (or even arbitrarily many inner problems) may be solved separately because a balance with the outer problem is not required. Now if a rigid body motion $\bU=\UD\be_x$ (due to the inner problem) is added to the surface deformations of the outer problem it has the sole effect of adding a plug flow solution to the swimming problem; in the zero Reynolds number limit a rigid body motion of two dimensional surface diffuses to infinity instantaneously. An interesting consequence of this is that when there arises a nonzero relative velocity, the idea of collective motion loses meaning, even for identical sheets, as in the frame moving with lower sheet (see Fig.~\ref{system}) we find different values for the flow at infinity, $U$ when $y\rightarrow -\infty$ and $U+\UD$ when $y\rightarrow \infty$. Further, if the sheets are different, then even if the inner problem demands $\UD=0$, the outer problem on either side produces flows at infinity (in the frame moving with the sheets) which are distinct. However even in the three dimensional case, when the swimmers are the same, synchronization is clearly driven by the forces between the bodies and those forces will dominate when the cells are close; because of this we expect to garner the leading order behavior from analysis of the inner problem.

\subsection{Avenues for future work}

In the problem considered in this paper, we have prescribed a front-back asymmetry in the waveforms propagated in our model of flagellated cells in order to give rise to synchronization. Real eukaryotic  flagella deform under applied (internal) forces, and this deformation may provide an additional mechanism of symmetry-breaking. Indeed recent experiments using rotating paddles suggest that elastic deformation is a key requirement to obtain synchrony for a geometry that is otherwise too symmetric to yield stable fixed points \cite{qian09}. Flagella flexibility  might thus be an important physical ingredient in the synchronizing dynamics, and we will address its relevance in future work.

\begin{acknowledgments}
We thank David Woolley for allowing us to reproduce the experimental picture from Ref.~\cite{woolley09}, as well as Lisa Fauci and Saverio Spagnolie for useful discussions.  Funding by the NSF (CBET-0746285) and NSERC (PGS D3-374202) is gratefully acknowledged.
\end{acknowledgments}

\bibliography{synchronization}

\end{document}